\documentclass[%
reprint,
superscriptaddress,
 amsmath,amssymb,
 aps,
pre,
floatfix,
]{revtex4-1}
\usepackage[utf8]{inputenc}
\usepackage{graphicx}
\usepackage{dcolumn}
\usepackage{bm}
\usepackage[caption=false]{subfig}
\usepackage{color}

\begin{document}

\preprint{APS/123-QED}

\title{Semi-Lagrangian lattice Boltzmann method for compressible flows}
\author{Dominik Wilde}
\email{wilde.aerospace@gmail.com}
\affiliation{
 Department of Mechanical Engineering, University of Siegen, Paul-Bonatz-Straße 9-11, 57076 Siegen-Weidenau, Germany}%
\affiliation{
Institute of Technology, Resource and Energy-efficient Engineering (TREE), Bonn-Rhein-Sieg University of Applied Sciences,
Grantham-Allee 20, 53757 Sankt Augustin, Germany
}

\author{Andreas Krämer}%
 \affiliation{
 National Heart, Lung, and Blood Institute, National Institutes of Health, Bethesda, MD 20892, United States
}%
 \author{Dirk Reith}
\affiliation{
Institute of Technology, Resource and Energy-efficient Engineering (TREE), Bonn-Rhein-Sieg University of Applied Sciences,
Grantham-Allee 20, 53757 Sankt Augustin, Germany
}
\affiliation{Fraunhofer Institute for Algorithms and Scientific Computing (SCAI), Schloss Birlinghoven, 53754 Sankt Augustin, Germany}

\author{Holger Foysi}

\affiliation{
 Department of Mechanical Engineering, University of Siegen, Paul-Bonatz-Straße 9-11, 57076 Siegen-Weidenau, Germany}%

\date{\today}

\begin{abstract}

This work thoroughly investigates a semi-Lagrangian lattice Boltzmann (SLLBM) solver for compressible flows. In contrast to other LBM for compressible flows, the vertices are organized in cells, and interpolation polynomials up to fourth order are used to attain the off-vertex distribution function values. Differing from the recently introduced Particles on Demand (PoD) method \cite{Dorschner2018}, the method operates in a static, non-moving reference frame. Yet the SLLBM in the present formulation grants supersonic flows and exhibits a high degree of Galilean invariance.
The SLLBM solver allows for an independent time step size due to the integration along characteristics and for the use of unusual velocity sets, like the D2Q25, which is constructed by the roots of the fifth-order Hermite polynomial. 
The properties of the present model are shown in diverse example simulations of a two-dimensional Taylor-Green vortex, a Sod shock tube, a two-dimensional Riemann problem and a shock-vortex interaction. It is shown that the cell-based interpolation and the use of Gauss-Lobatto-Chebyshev support points allow for spatially high-order solutions and minimize the mass loss caused by the interpolation. Transformed grids in the shock-vortex interaction show the general applicability to non-uniform grids.
\end{abstract}

\pacs{Valid PACS appear here}
\maketitle


\section{\label{sec:intro}Introduction}
In the field of weakly compressible and isothermal flows, the lattice Boltzmann method (LBM) \cite{McNamara1988,Higuera1989,Chen1991} has emerged as an efficient numerical solver that suits modern, highly-parallel computing architectures. 
Consequently, many attempts have been undertaken to extend the method to energy conserving flows \cite{Alexander1992}. Despite considerable progress in this field, the research on robust and efficient lattice Boltzmann models for compressible flows is still ongoing. 

Existing approaches can generally be categorized in two ways \cite{Kruger2016}: on- versus off-lattice, and single- versus double-population. While on-lattice approaches inherit the LBM's exact, space-filling streaming step \cite{Frapolli2015,Atif2018,Saadat2019, Latt2019}, off-lattice methods discretize the Boltzmann equation through finite volume \cite{Chen1998, Zhang2001, Guo2015,Feng2016} or finite difference schemes \cite{Shi2001,WANG2007,Nie2009}. 
Orthogonal to this classification, single-population models \cite{Alexander1992, Frapolli2017} express all physical moments, including energy and heat flux, by a single distribution function. In order to represent all moments, the particle velocity sets have to be extended beyond the lattices typically used for weakly compressible flows. 
In contrast, double-population models \cite{Kataoka2004,Li2007,Saadat2019} represent the local internal energy through a separate distribution function that is coupled with the density and momentum coming from the first distribution function. 

The works of Frapolli \cite{Frapolli2017} and Coreixas \cite{Coreixas2018} present an excellent overview of compressible extensions of the LBM. They show how previous approaches have traditionally suffered from instability, numerical dissipation, large computational cost, and undesirable couplings between the physical and numerical parameters.


The use of an efficient off-lattice method is key, since compressible on-lattice Boltzmann methods suffer from decisive disadvantages. 
First, the time step size is set to unity and cannot be changed independently from the physical parameters. In the weakly compressible regime, the Mach number is merely a numerical parameter following the law of similarity, so that the time step size can be changed via the flow velocity \cite{Kruger2016}. In compressible flows, however, the Mach number regains its meaning as a physical quantity with only limited remaining options to change the time step, e.g by increasing or decreasing the spatial resolution, which is still coupled to the time step size. Alternatively, the speed of sound can be varied, at the price of a inferior numerical stability \cite{Alexander1992}.
Second, the simulation of energy conserving flows requires disproportionately large velocity sets since integer-based values of the particle velocities lead to leading-order errors in the relevant moments of the Maxwell-Boltzmann distribution function \cite{Chikatamarla2009}. A common way to reduce these errors is to increase the number of abscissae, leading to velocity sets like D1Q7, D2Q49 or D3Q343 \cite{Frapolli2016}. For weakly compressible flows, the D1Q3 velocity set and its tensor products D2Q9 and D3Q27 in combination with a second-order truncation of the Maxwell-Boltzmann distribution mostly match the desired properties of the Navier-Stokes equations in the hydrodynamic limit, except the "famous $\mathcal{O}(Ma^3)$ error term" \cite{Coreixas2017}, which breaks Galilean invariance and restricts simulations to low Mach numbers.

The physical and numerical parameters can be decoupled in an off-lattice framework, where the velocity set does not need to match the computational grid.
In the field of off-lattice Boltzmann methods the semi-Lagrangian lattice Boltzmann method (SLLBM) was recently introduced \cite{Kramer2017a} and further investigated in a number of subsequent works \cite{Kramer,DiIlio2018,Dorschner2018,Kramer2018,Zarikov2019}. To formulate the streaming step the algorithm follows the trajectories of the lattice Boltzmann equation along their characteristics to find the cells of the corresponding departure points. In each cell, a finite element formulation reconstructs the distribution function values from the local degrees of freedom by interpolation polynomials.
The SLLBM overcomes the aforementioned drawbacks of the standard lattice Boltzmann method. It decouples the time step from the grid spacing and allows for high-order spatial accuracy on irregular grids.  In common with Eulerian-based off-lattice Boltzmann solvers the SLLBM enables the use of non-integer-based velocity sets. Yet, compared to Eulerian-based off-lattice Boltzmann schemes, the streaming step is computationally more efficient \cite{Kramer2018}, reduces numerical dissipation, and is still close to the original formulation of the LBM, as it also follows the trajectories of the discrete Boltzmann equation back in time \cite{Kramer2017a}. The problems that arise with interpolation-based schemes have been discussed in earlier works, e.g. by Chen \cite{Chen1998}, like the issue that interpolation-based LBM do not conserve mass, momentum and energy. However, this effect can be at least reduced by using high-order interpolation polynomials, which we showed in a recent work \cite{Kramer2018}. 

The present work thoroughly investigates an SLLBM solver for compressible flows. The methodology is related to the Particles on Demand (PoD) method \cite{Dorschner2018}, which attained considerable interest \cite{Zarikov2019}. The PoD made use of a semi-Lagrangian streaming step to attain the off-lattice distribution function values in a dynamically shifting reference frame, similar to other adaptive LBM models for compressible flows \cite{Sun1998,Sun2000,Sun2003}.  In contrast to PoD, we restrict the present method to static, non-moving reference frames at a fixed reference temperature and show that the method still remains competitive in terms of Galilean invariance up to a certain Mach number. In that sense, the present method is closer to the original lattice Boltzmann method formulation, which also operates in a static, non-moving reference frame.

In addition to the achievements of PoD, we use Hermite-based equilibria \cite{Shan1998,Shan2006} and high-order interpolation polynomials. We apply the double-population off-lattice approach that mitigates the aforementioned issues of previous solvers and provides a more comprehensive solution for simulating compressible flows with the LBM. 
 
The semi-Lagrangian streaming step also offers more flexibility in constructing velocity sets from Hermite polynomials \cite{Bardow2008,Shan2016}, which is particularly relevant for compressible flow.
Originating in Grad's work on the moment system in kinetic theory \cite{Grad1949}, a number of subsequent works showed that consistent equilibrium distribution functions can be derived by projecting the Boltzmann equation onto a low-dimensional Hilbert subspace through Hermite expansion \cite{Shan2006,Chikatamarla2009,Shan2016}. This method is used to construct high-order velocity sets that recover the compressible Fourier-Navier-Stokes equations in the hydrodynamic limit \cite{Malaspinas2009,Coreixas2017}.\\[.3cm]
This article is structured as follows. Section \ref{sec:methodology} derives the equilibria and velocity sets and recapitulates the semi-Lagrangian advection scheme. Section \ref{sec:results} validates the method by simulations of a moving incompressible Taylor-Green vortex, a smooth density propagation, a Sod shock tube, a two-dimensional Riemann problem, and a shock-vortex interaction. Section \ref{sec:discussion} discusses the properties of the compressible SLLBM and highlights the advantages over other compressible LBM solvers. Section \ref{sec:conclusion} provides conclusions.

\section{Methodology}
\label{sec:methodology}
The Boltzmann equation reads
\begin{equation}
    \partial_t f + \boldsymbol{\xi} \cdot \boldsymbol{\nabla} f = \Omega (f)
\end{equation}
with the particle distribution $f$, the particle velocities $\mathbf{\xi}$ and the collision operator $\Omega$. In this work, only the linearized single relaxation time collision operator proposed by Bhatnagar-Gross-Krook (BGK) \cite{Bhatnagar1954} will be used:\vspace{-.3cm}
\begin{equation}
\Omega (f) = - \frac{1}{\lambda} \left(f - f^{\mathrm{eq}}\right),
\end{equation}
where $f^{\mathrm{eq}}$ denotes the equilibrium distribution function, while the relaxation time $\lambda = \nu / c_s^2$ reflects the kinematic viscosity $\nu$ and the speed of sound $c_s$.
The macroscopic density ($\rho$), velocity ($u$), and energy ($E$) are represented by the moments of the distribution function: \vspace{-.3cm}
\begin{align}
    \rho &= \int f d \boldsymbol\xi \\
    \rho \boldsymbol u &= \int f \boldsymbol\xi d \boldsymbol\xi \\
    2 \rho E &= \int (f |\boldsymbol\xi|^2 + g) d \boldsymbol\xi, \label{eq:E}
\end{align}
The second distribution function $g$ represents the rotational energy of the polyatomic molecules and is described in subsection \ref{sec:gamma}.
\subsection{Equilibrium distribution function}\label{sec:eq}
One key component to correctly calculate thermal and compressible flows is the construction of the equilibrium distribution function $f^\mathrm{eq}$. The present section is based on the comprehensive derivation of the equilibrium, which was detailed by Shan et al. \cite{Shan2006}. The equilibrium is found via scaled Hermite polynomials $\boldsymbol{\mathcal{H}}^{(n)}(\hat{\boldsymbol\xi}),$ which are formulated in terms of the normalized particle velocities $\hat{\boldsymbol\xi} = \boldsymbol\xi / c_s.$
The Hermite polynomials are defined by the generating function \vspace{-.3cm}
\begin{equation}
\omega\left(\hat{\boldsymbol\xi}\right) = \frac{1}{\sqrt{2\pi}} \mathrm{exp}\left(- \frac{|\hat{\boldsymbol{\xi}}|^2}{2}\right)
\end{equation}
\vspace{-.3cm}as\vspace{-.3cm}
\begin{equation}
\boldsymbol{\mathcal{H}}^{(n)}\left(\hat{\boldsymbol\xi}\right) = (-1)^n\frac{1}{\omega(\hat{\boldsymbol\xi})}\frac{d^n}{d\boldsymbol\xi^n}\omega(\hat{\boldsymbol\xi}),
\end{equation}
where $c_s^2 = R T_0$ is the product of gas constant and reference temperature. (Note that the generating function is differentiated with respect to $\boldsymbol\xi$ rather than $\hat{\boldsymbol\xi}.$)\\
The scaled Hermite polynomials up to fourth order are 

\begin{align*}
    \boldsymbol{\mathcal{H}}^{(0)}&= 1 \\
    \boldsymbol{\mathcal{H}}^{(1)}&=
                   \frac{\hat{\boldsymbol{\xi}}}{c_s}  \\
   \boldsymbol{\mathcal{H}}^{(2)}&= \frac{\hat{\boldsymbol{\xi}}^2-\boldsymbol{I}}{c_s^2} \\
    \boldsymbol{\mathcal{H}}^{(3)}&= \frac{\boldsymbol{\hat\xi}^3-3\overbrace{\hat{\boldsymbol{\xi}}\boldsymbol{I}}}{c_s^3} \\ 
    \boldsymbol{\mathcal{H}}^{(4)}&= \frac{\hat{\boldsymbol{\xi}}^4-6\overbrace{\hat{\boldsymbol{\xi}}^2\boldsymbol{I}}+3\overbrace{\boldsymbol{I}\boldsymbol{\delta}}}{c_s^4}.
\end{align*}
Here, $\boldsymbol{I}$ is the identity matrix, $\boldsymbol\xi^n:=\boldsymbol\xi\otimes\boldsymbol\xi\otimes \cdots \otimes \boldsymbol\xi$ ($n$-times) and for arbitrary vectors, for example, we write $\boldsymbol a\boldsymbol b:=\boldsymbol a\otimes\boldsymbol b$. Furthermore, we defined for a $n$-th order tensor $\boldsymbol T^{(n)}$, according to \cite{Suchy1996}, the relation
\[\overbrace{\boldsymbol T^{(n)}}=\frac{1}{number~ of~ \pi}\sum_\pi \boldsymbol T^{(n)},\]
with $\pi$ indicating the significant permutations of either the base vectors or indices. For arbitrary vectors $\boldsymbol v$ and $\hat{\boldsymbol\xi}$ one has, for example,
\[\overbrace{\boldsymbol{v}\hat{\boldsymbol\xi}^2}=\frac{1}{3}(
\boldsymbol{v}\hat{\boldsymbol\xi}\hat{\boldsymbol\xi}+
\hat{\boldsymbol\xi}\boldsymbol{v}\hat{\boldsymbol\xi}+
\hat{\boldsymbol\xi}\hat{\boldsymbol\xi}\boldsymbol{v}).
\]
\\[-0.2cm]
The (equilibrium) distribution functions can be expressed as a Hermite series \cite{Coreixas2017} of order $N$
\begin{eqnarray}\label{eq:nonequilibrium}
   f^{N}(\mathbf{x},\hat{\boldsymbol{\xi}},t) &\approx&  \omega(\hat{\boldsymbol{\xi}}) \sum_{n=0}^N \frac{1}{n!}\boldsymbol{a}^{(n)}(\mathbf{x},t) :  \boldsymbol{\mathcal{H}} ^{(n)}(\hat{\xi}),\\ \label{eq:equilibrium} f^{\mathrm{eq},N}(\mathbf{x},\hat{\boldsymbol{\xi}},t) &\approx&  \omega(\hat{\boldsymbol{\xi}}) \sum_{n=0}^N \frac{1}{n!}\boldsymbol{a}^{(n)}_{eq}(\mathbf{x},t) :  \boldsymbol{\mathcal{H}} ^{(n)}(\hat{\xi}),
\end{eqnarray}
with $:$ indicating full contraction. The order $N$ determines the physics of the lattice Boltzmann model. 
The Hermite coefficients $\boldsymbol{a}$ are obtained via projection of $f$ on the orthogonal basis of Hermite tensors. 
The equilibrium coefficients $\boldsymbol{a}^{(n)}_{eq}$ are therefore
directly related to the moments of the Maxwell-Boltzmann equilibrium, as \cite{Coreixas2017}
\begin{subequations}
\label{moments}
\begin{eqnarray}
    a^{(0)}_{eq} &=& \rho \\
    a^{(1)}_{\alpha,eq} &=& \rho u_\alpha \\ 
    a^{(2)}_{\alpha \beta,eq} &=& \Pi_{\alpha \beta}^\mathrm{eq} = \rho (u_\alpha u_\beta + T_0(\theta -1 )\delta_{\alpha\beta}) \\
    a^{(3)}_{\alpha \beta \gamma,eq} &=& \mathcal{Q}_{\alpha \beta \gamma}^\mathrm{eq} = \rho \left[u_\alpha u_\beta u_\gamma +  T_0(\theta -1 )(\delta_{\alpha\beta}u_\gamma  \right.\nonumber\\
     & & \left.\hspace{1.2cm} + \delta_{\alpha\gamma}u_\beta +\delta_{\beta\gamma}u_\alpha)\right] \label{moments_Q}\\
    a^{(4)}_{\alpha \beta \gamma \delta,eq} &=& \mathcal{R}_{\alpha \beta \gamma \delta}^\mathrm{eq} \nonumber\\
    &=& \rho[ u_\alpha u_\beta u_\gamma u_\delta   
     + T_0(\theta-1)((\delta_{\alpha\beta}\delta_{\gamma\delta} \nonumber\\
     & &+ \delta_{\alpha\gamma}\delta_{\beta\delta}+ \delta_{\alpha\delta}\delta_{\beta\gamma} )T_0(\theta-1)  \nonumber\\ 
   & & + \delta_{\alpha\beta} u_{\gamma}u_{\delta} + \delta_{\alpha\gamma} u_{\beta}u_{\delta} + \delta_{\alpha\delta} u_{\beta}u_{\gamma} +\nonumber\\
   & & \delta_{\beta\gamma} u_{\alpha}u_{\delta} + \delta_{\beta\delta} u_{\alpha}u_{\gamma} + \delta_{\gamma\delta} u_{\alpha}u_{\beta}  )] 
\vspace{.3cm}
\end{eqnarray}
\end{subequations}
Note that in case of the established weakly compressible lattice Boltzmann simulations, the isothermal assumption leads to a local temperature of $\theta=T/T_0=1$ and to vanishing terms in Eqs. \eqref{moments}.

Truncating at second order covers the well-known lattice Boltzmann model for weakly compressible flows at low velocity. Truncating at third order compromises the non-equilibrium parts of $\mathcal{Q}_{\alpha\beta\gamma}$ due to deviations in $\mathcal{R}_{\alpha\beta\gamma\delta}^\mathrm{eq}.$ Thus, to represent the heat flux correctly, a fourth-order expansion is required \cite{Shan2006,Coreixas2017}. 
Consequently, to recover the full Fourier-Navier-Stokes equations, the approximation in equations \eqref{eq:nonequilibrium} and \eqref{eq:equilibrium} is of fourth order in the present work. 

\subsection{Hermite polynomial velocity sets}\label{sec:velocity_set}

\begin{figure}[htp]
    \centering
    \includegraphics[width=0.5\linewidth]{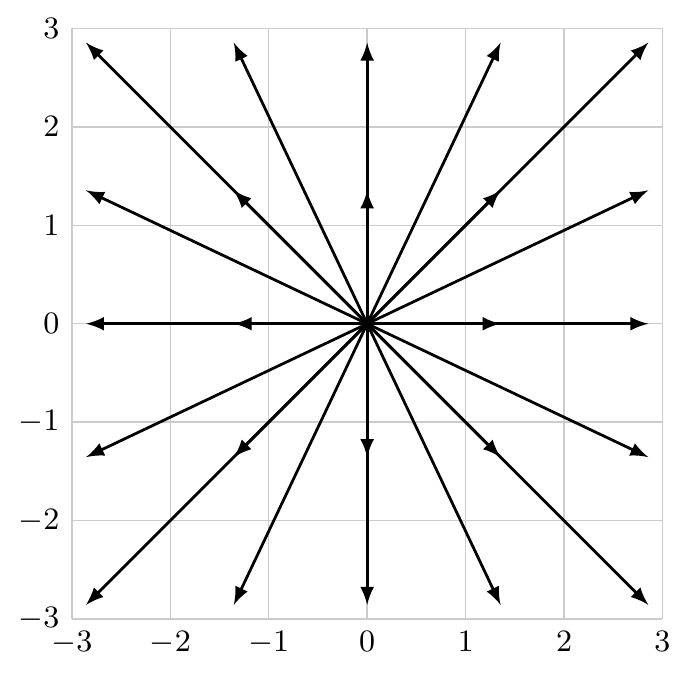}
    \caption{D2Q25 velocity set based on Hermite polynomials of order 5. Although the velocity can be scaled, matching the regular grid is not possible.}
    \label{fig:stencil}
\end{figure}

To recover the $M$th order moment $\boldsymbol{a}^M$, the velocity space is discretized by a (weighted) Gauss-Hermite quadrature 
\begin{eqnarray}
        \boldsymbol{a}^M =  \int  f^{N} \boldsymbol{\mathcal{H}}^M d\hat{\boldsymbol{\xi}} &=& \sum_{i=1}^V f_i^{N} \boldsymbol{\mathcal{H}}^M(\hat{\boldsymbol{\xi}}_i),\\
    \boldsymbol{a}^M_{eq}  =  \int  f^{\mathrm{eq},N} \boldsymbol{\mathcal{H}}^M d\hat{\boldsymbol{\xi}} &=& \sum_{i=1}^V f_i^{\mathrm{eq},N} \boldsymbol{\mathcal{H}}^M(\hat{\boldsymbol{\xi}}_i),
    \label{discrete}
\end{eqnarray}
provided $N \geq M$ and with the discrete equilibrium
\begin{equation*}
    f^{\mathrm{eq},N}_i = {w}_i/w(\hat{\boldsymbol{\xi}}_i) \cdot f^{\mathrm{eq},N}(\hat{\boldsymbol{\xi}}_i).
\end{equation*}
Here, $w_i$ represents the weights coming from a Gauss-Hermite quadrature, $V$ is the number of velocities in the velocity set. In general, the number of possible velocity sets is several times larger than in on-lattice Boltzmann methods, since the abscissae of the velocity sets are not required to match the grid points. The present work uses a D2Q25 velocity set, which is derived by a tensor multiplication from the one-dimensional D1Q5 velocity set \cite{Shan2016}. The one-dimensional abscissae and weights are listed in Table \ref{tab:d1q5}, while the two-dimensional D2Q25 velocity set is depicted in Fig. \ref{fig:stencil}. Unless scaled, the characteristic lattice speed of sound equals $c_s=1$.
\begin{table}
    \centering
    \caption{Abscissae $\xi_i$ and weights $w_i$ for the D1Q5 lattice. The D2Q25 velocity set is obtained by a tensor multiplication.}
    \label{tab:d1q5}

\renewcommand{\arraystretch}{1.5}
\setlength{\tabcolsep}{7pt}
\begin{tabular}{ c c c } 
  $i$ & $\xi_i$ & $w_i$\\
  \hline
  $0$ & $0$ & $8/15$ \\
  $1,3$ & $\pm\sqrt{5 - \sqrt{10}}$ & $(7+2\sqrt{10})/60$ \\
  $2,4$ & $\pm\sqrt{5 + \sqrt{10}}$  & $(7-2\sqrt{10})/60$ \\

\end{tabular}
\end{table}

This off-lattice velocity set is of quadrature order $Q=9$ and consequently recovers moments up to order $N=4$, fulfilling the requirement $(M+N)/2\leq Q$ to recover the Fourier-Navier-Stokes equations \cite{Nie2008a}.

\subsection{Variable Prandtl number}\label{sec:prandtl}
In the context of thermal and compressible flows, using the BGK operator limits the Prandtl number $Pr$, i.e. the ratio of kinematic viscosity $\nu$ and thermal diffusivity $\mathcal{\alpha}$, to $Pr = \nu /\mathcal{\alpha} = 1$.  This drawback can be circumvented by using a multi-relaxation time collision operator to decouple both properties of the fluid \cite{Shan2007}, a quasi-equilibrium \cite{Ansumali2007,Frapolli2015} or by using a BGK-Shakhov model \cite{Shakhov1972}. Although we observed that for the test cases in this work there is no significant difference, the SLLBM solver is capable to adapt the Prandtl number by the quasiequilibrium $f_i^*$ and $g_i^*$ \cite{Ansumali2007,Frapolli2015}, which is

\begin{equation}
    f^*_i = w_i \bar{\mathbf{Q}}^\mathrm{neq} : ( \hat{\boldsymbol{\xi}_i^3} - 3 \overbrace{ \hat{\boldsymbol{\xi}_i} \boldsymbol{I}})/(c_s^3),
\end{equation}

with the non-equilibrium part of the centered heat flux tensor $\bar{\mathbf{Q}}^\mathrm{neq} = \bar{\mathbf{Q}} - \bar{\mathbf{Q}}^\mathrm{eq} $

\begin{equation}
    \bar{\mathbf{Q}} = \sum_{i=1}^n f_i (\hat{\boldsymbol{\xi}_i}-\boldsymbol{u}) \otimes( \hat{\boldsymbol{\xi}_i}-\boldsymbol{u}) \otimes( \hat{\boldsymbol{\xi}_i}-\boldsymbol{u}),
\end{equation}

with the same proceeding for $g_i^*$.

\subsection{Variable heat capacity ratio}\label{sec:gamma}
As the Boltzmann equation and the Maxwell-Boltzmann distribution are valid in the assumption of monoatomic flows, the heat capacity ratio $\gamma = C_p / C_v $ is fixed to $\gamma = (D+2)/2$, with the dimension $D$ and the heat capacities at constant pressure and volume $C_p$ and $C_v$, respectively. Nonetheless, several attempts in the literature solve this issue by using a second set of distribution functions \cite{Nie2008,Frapolli2017}, inspired by the Rykov model in kinetic theory for polyatomic gases \cite{Rykov1976}, or by adapting the equilibrium distribution to the heat capacity of the fluid \cite{Dellar2008}. 
We adopt the ansatz by Frapolli et al. for the LBM by incorporating an additional set of distributions $g$ and its equilibrium $g^\mathrm{eq}$.

The equilibrium of $g^\mathrm{eq}$ can easily be determined as 
\begin{equation}
    g_i^\mathrm{eq} = (2 C_v - D) \theta f_i^\mathrm{eq}
\end{equation}
with $\gamma=R/C_v+1$, where the gas constant $R$ is usually set to unity (without loss of generality). The distribution function $g_i$ follows the same stream and collide algorithm as the distribution function $f_i$ and is ultimately used in equation (\ref{eq:E}).

The lattice Boltzmann equations of the present model read
\begin{widetext}
\begin{subequations}
\label{eq:streaming_fg}
\begin{align} 
    	f_{i}(x, t) &= 
		f_{i}(x-\delta_{t} \xi_{i}, t-\delta_{t}) - \frac{1}{\tau} 
		\left[ f_{i}(x-\delta_{t} \xi_{i}, t-\delta_{t})- f_{i}^{\mathrm{eq}}(x-\delta_{t} \xi_{i}, t-\delta_{t}) \right] + \left(\frac{1}{\tau}-\frac{1}{\tau_\mathrm{Pr}}\right)f_i^*(x-\delta_{t} \xi_{i}, t-\delta_{t}) \\
		g_{i}(x,t) &= 
		g_{i}(x-\delta_{t} \xi_{i}, t-\delta_{t}) - \frac{1}{\tau} 
		\left[ g_{i}(x-\delta_{t} \xi_{i}, t-\delta_{t})- g_{i}^{\mathrm{eq}}(x-\delta_{t} \xi_{i}, t-\delta_{t}) \right] + \left(\frac{1}{\tau}-\frac{1}{\tau_\mathrm{Pr}}\right)g_i^*(x-\delta_{t} \xi_{i}, t-\delta_{t}),
\end{align}
\end{subequations}
\end{widetext}
with the relaxation time $\tau = \nu / (c_s^2\delta_t) + 0.5$, where the term of $0.5$ is a consequence of the second-order time integration in terms of the trapezoidal rule \cite{He1998a,Krause2010,Wilde2019}. The relaxation time $\tau_\mathrm{Pr} = (\tau -0.5)/Pr + 0.5$ is related to the Prandtl number $Pr$. In contrast to the usual lattice Boltzmann algorithm, neither the time step size $\delta_t$, nor the expression $\delta_t\xi_i$ are required to be integers due to the use of a semi-Lagrangian streaming step.

\subsection{Semi-Lagrangian streaming step}\label{sec:sllbm}
To obtain the off-vertex distribution function values $f_i$ (and likewise $g_i$), the SLLBM uses a cell-wise polynomial interpolation. Therefore, the computational domain is divided into a finite number of cells $N_\Xi$ with Lagrange polynomials $\psi_\Xi$ of order $p$ as shape functions. The distribution function values at the support points in cell $\Xi$ are denoted as $\hat{f}_{i\Xi j},$ so that
\begin{equation}\label{eq:sl_streaming}
f_i(\mathbf{x},t) = \sum_{\Xi =1}^{N_{\Xi}} \sum_{j=1}^{N_{j}} \hat{f}_{i\Xi j}(t) \psi_{\Xi j}(\mathbf{x})
\end{equation}
for all points $\mathbf{x}$ in cell $\Xi.$

The distributions $\hat{f}_{i\Xi j}$ are updated via the Equations \eqref{eq:streaming_fg}, where the value at the departure point $f_i(x-\delta_t \xi_i)$ is obtained via the interpolation defined in Eq. \eqref{eq:sl_streaming}.
An equidistant interpolation is stable for low order interpolation polynomials only. Therefore, at higher interpolation orders this work applies Gauß-Lobatto-Chebyshev support points to dampen the Runge phenomenon and to increase the stability. \cite{Kramer2017a,Hesthaven2002}. 
Gauß-Lobatto-Chebyshev support points in one dimension read

\begin{equation*}
    x_k = \frac{1}{2}\left[-\cos\left(\frac{k-1}{n-1} \pi \right)+ 1 \right], \ k = 1,2,\ldots,n.
\end{equation*}

\begin{figure}
    \centering
    \includegraphics[width=0.4\linewidth]{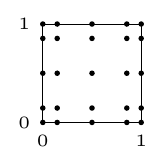}
    \caption{Distribution of Gauß-Lobatto-Chebyshev support points on a two-dimensional reference cell with $p=4$.}
    \label{fig:reference_cell}
\end{figure}

The two-dimensional unit cell used in this work is depicted in Fig. \ref{fig:reference_cell}.

The interpolation error of semi-Lagrangian advection schemes has been investigated, for instance, by Einkemmer and Ostermann \cite{Einkemmer2015} or Falcone and Ferretti \cite{Falcone2013}. For the SLLBM used in the present work, we have previously discussed the nontrivial behavior of the temporal discretization error \cite{Kramer2018}, which is of order 
\begin{equation}
\mathcal{O}\left(\mathrm{min}\left(\frac{\delta_x^{p+1}}{\delta_t}, \delta_x^p \right) + \delta_t^2\right).
\end{equation}
This means in practice that the error has a minimum for an intermediate CFL numbers and is bounded by the spatial discretization error in the limit of small time steps.

All SLLBM simulations in this manuscript have been performed with the \emph{NATriuM} library \cite{Kramer2018}, which makes use of the finite element library \emph{deal.ii} \cite{Bangerth2007}. See \cite{Kramer2018} for further details concerning the implementation.

\subsection{Thermo-hydrodynamic limit}

The thermo-hydrodynamic limit of the presented equilibrium distribution function 
was presented and thoroughly discussed by Malaspinas et al. \cite{Malaspinas2009} and by Coreixas et al. \cite{Coreixas2017}, who found
\begin{align}
    \partial_{t}(\rho) +\partial_\alpha (\rho u_\alpha) &= 0 \\
    \partial_{t}(\rho u_\alpha) + \partial_\beta (\rho u_\alpha u_\beta) &= \partial_\alpha \sigma_{\alpha \beta} \nonumber \\ 
    \partial_t(\rho E ) + \partial_\alpha (\rho E u_\alpha) &= \partial_\alpha (\kappa \partial_\alpha T) + \partial_\alpha (\sigma_{\alpha\beta} u_\beta) 
\end{align}
with the stress tensor
\begin{equation}
    \sigma_{\alpha \beta} = - P \delta_{\alpha \beta} \label{eq:momentum} + \nu \partial_\beta \left[ \rho \left(\partial_\beta u_\alpha + \partial_\alpha u_\beta \right) \right] - \zeta \partial_\gamma u_\gamma \delta_{\alpha\beta},
\end{equation}

with the kinematic viscosity 
\begin{equation}
    \nu = \left(\tau-\frac{1}{2}\right) c_s^2 \theta \delta_t,
\end{equation}

and with the heat conductivity 

\begin{equation}
    \kappa = \left(\tau_\mathrm{Pr}-\frac{1}{2}\right) C_p c_s^2 \theta \delta_t.
\end{equation}

The bulk viscosity equals \cite{Frapolli2016}

\begin{equation}
    \zeta = \rho \nu \left(\frac{1}{C_v} -\frac{2}{D}\right),
\end{equation}
with $C_v$ and $C_p$ being the heat capacity at constant volume and pressure, respectively.

\subsection{On-lattice Boltzmann solver for comparison}\label{sec:on-lattice}
Section \ref{sec:shocktube} compares the SLLBM with an on-lattice Boltzmann solver. For the latter, we chose the D2V37 velocity set proposed by Philippi et al. \cite{Philippi2006}, which is of the same quadrature order $Q=9$ as the D2Q25 in this work. Additionally, the equilibrium in Subsection \ref{sec:eq} and the second distribution function in Subsection \ref{sec:gamma} were used to obtain a variable adiabatic exponent $\gamma$.
The on-lattice simulations were performed by the solver \emph{Lettuce} \cite{lettuce}, which is based on the machine learning framework \emph{PyTorch} \cite{pytorch}. The code is written in Python and makes use of the PyTorch routines to enable GPU simulations.

\section{Results}
\label{sec:results}
In this section, a variety of test cases will be studied. First, the errors of the method are quantified by simulations of the two-dimensional incompressible Taylor-Green vortex. Then the accuracy will be confirmed by a smooth density propagation.  The Sod shock tube shows the general capability of the SLLBM to deal with shocks at the original density ratio presented by Sod \cite{Sod1978}. Then the two-dimensional Riemann problem will be studied, followed by a simulation of a shock-vortex interaction on non-uniform grids.

\subsection{Moving two-dimensional Taylor-Green vortex} \label{sec:tgv2d}

The SLLBM was validated by a moving incompressible two-dimensional Taylor-Green vortex. As it comes along with a reference solution it is suited to quantify the interpolation-caused errors and to show the high degree of Galilean invariance if an appropriate velocity set is used. The initial and reference solution is
\begin{eqnarray*}
u_x^\mathrm{ref}(x,y,t)&=&+u_v\sin(x-u_h t)\cos(y)\mathrm{exp}(-2\nu t) \\
u_y^\mathrm{ref}(x,y,t)&=&-u_v\cos(x-u_h t)\sin(y)\mathrm{exp}(-2\nu t) \\
P^\mathrm{ref}(x,y,t)&=&\frac{1}{4}\!\left(\cos(2(x\!-\!u_h t))\!+\!\cos(2y))\mathrm{exp}(-4\nu t\right)\!.
\end{eqnarray*}

with the constant horizontal velocity $u_h$ and the vortex velocity $u_v$. The Reynolds number was set to $Re={u_v l}/{\nu}=10$, with an initial maximum speed $u_v=1$ on a domain of $l=2\pi$. All units were transferred to lattice units; the initial Mach number of the vortex flow was constantly $\mathrm{Ma_v}=0.01$, while the horizontal Mach number was either disabled for case (a) or set to $\mathrm{Ma_h}=0.05$ for case (b). The results were compared to the reference at $t=1.83$, corresponding to a decay of the vortex velocity to $u_\mathrm{max} = 0.1 u_v $. The grid consisted of 4, 8, 16, and 32 cells per direction times the interpolation order $p=4$, yielding  16, 32, 64, and 128 grid points per direction. The time step was set to $\delta_t=0.0008$ for the coarsest test case and halved for each refinement step, without changing the Mach numbers.

We compare two velocity sets: the standard D2Q9 velocity set and the D2Q25 velocity set, both with the SLLBM advection step. The fourth order equilibrium specified in Section \ref{sec:eq} was used in an isothermal configuration, i.e. $\theta=1$.

\begin{figure}
    \centering
    \includegraphics[width=\linewidth]{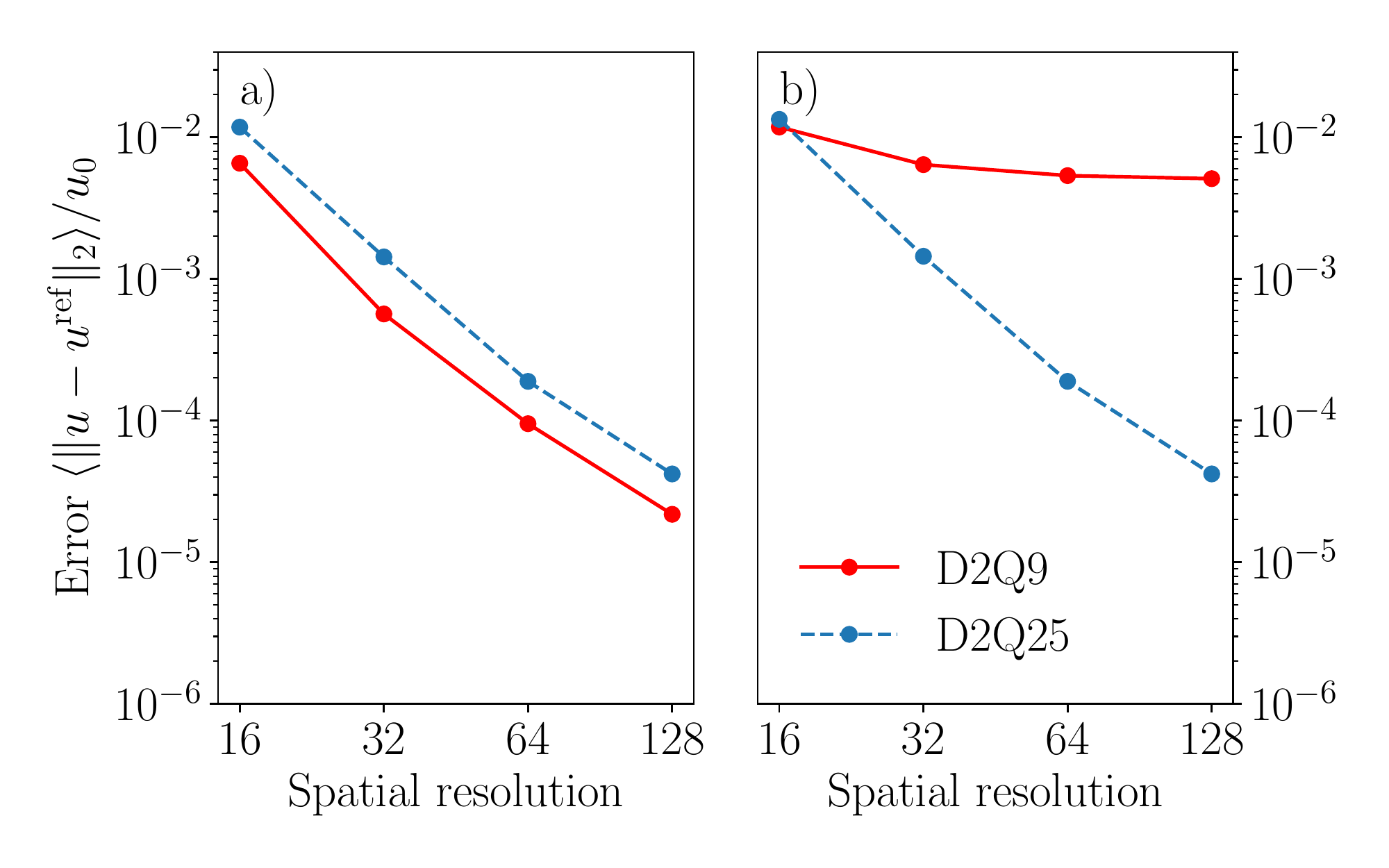}
    \caption{Numerical errors of incompressible 2D Taylor-Green vortex benchmarks with horizontal Mach numbers ${Ma_h=0}$~(a) and $Ma_h=0.05$~(b).}
    \label{fig:tgv}
\end{figure}

Figure \ref{fig:tgv} illustrates the errors of the two-dimensional Taylor-Green vortex flow. Subfigure (a) shows the non-moving case. As detailed in section \ref{sec:sllbm} the interpolation provoked non-negligible errors, which were in total two to three times larger for the D2Q25 velocity set in comparison to the D2Q9, caused by the different amount of interpolations needed to perform one time step. 
However, the larger velocity set pays off for constantly moving vortices in case (b). Here the D2Q9 simulation converged towards an error, which is caused by the lacking Galilean invariance of this particular velocity set. In contrast, the D2Q25's errors were nearly unchanged compared to simulations (a) with $u_h=0$.

\begin{figure}
    \centering
    \includegraphics[width=\linewidth]{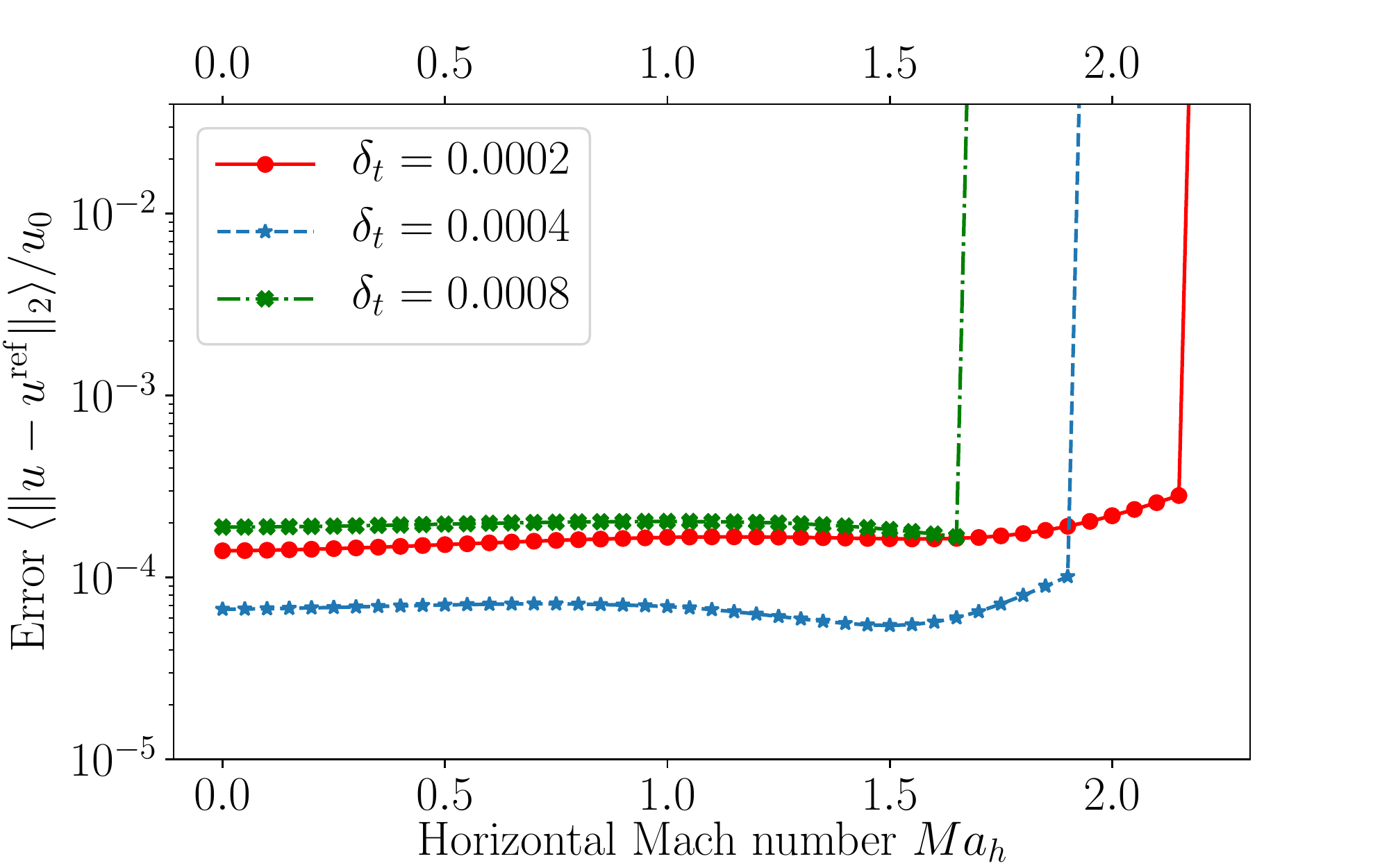}
    \vspace{-.5cm}
    \caption{Numerical errors of incompressible 2D Taylor-Green vortex benchmarks over horizontal Mach numbers $Ma_h$ with the D2Q25 velocity set for three different time step sizes.}
    \label{fig:maxMa}
\end{figure}

This test case also served to evaluate the maximum Mach number ranges of the D2Q25 velocity set. To achieve this, the horizontal velocity $Ma_h$ was increased until the simulations became unstable, while the initial vortex velocity remained at $Ma_v=0.01$. Fig. \ref{fig:maxMa} depicts the numerical errors over the horizontal Mach number $Ma_h$ for three different time step sizes with two main findings. First, there is a local minimum of the numerical error in case of the time step size $\delta_t=0.0004$. Second, for all three time step sizes the numerical errors remain more or less constant over the range of horizontal Mach numbers, while the Mach number limit was abruptly exceeded. Obviously, the maximum Mach number of semi-Lagrangian schemes is not solely determined by equilibrium and velocity set, but also by the time step size.

\begin{figure*}
    \centering
    \includegraphics[width=0.9\linewidth]{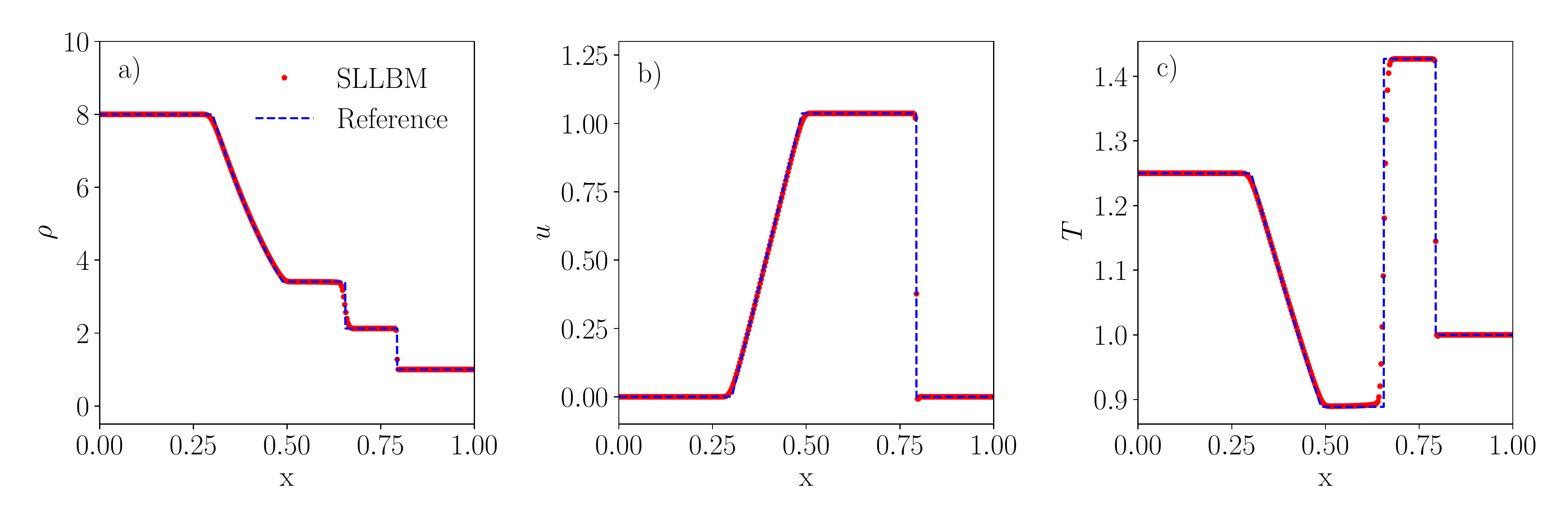}
    \vspace{-.5cm}
    \caption{Sod shock tube with $p=4, N = 800~\mathrm{(200~ cells)}, \delta t = 0.002$ after 735 iterations showing a) density $\rho$ , b) velocity $u$ and c) temperature $T$. The SLLBM is compared with Sod's reference results \cite{Sod1978}.}
    \label{fig:shocktubep4}
\end{figure*}

Fig. \ref{fig:shocktubep4} depicts the density, velocity, and temperature for the polynomial order $p=4$ at $t=0.15$, which is reached after 735 time steps. 

\subsection{Smooth density propagation}

To measure the accuracy of the SLLBM for compressible flows, the smooth density propagation \cite{Ji2018,Saadat2020} was evaluated. 

The exact solution is
\begin{eqnarray}
\rho(x,y,t) &=& 1 + 0.2\sin(\pi (x - u_x t)) \sin(\pi y) \\
u_x (x,y,t) &=& 1, u_y(x,y,t)=0, P(x,y,t) = 1
\end{eqnarray}

The domain was $x,y \in [-1,1]$ and the simulation end time was $t=2$. For the coarsest grid of $16\times16$ grid points (i.e. $4 \times 4$ cells and $p=4$), this time was reached in 1105 steps; the number of steps doubled for each finer simulation. The kinematic viscosity was set to $\nu = 10^{-10}$; the Mach number was set to $Ma=0.2$. 

Table \ref{tab:sdp} depicts the numerical errors of the density for six different resolutions. At coarse resolutions the numerical errors decreased with approximately fourth order of magnitude. At the finest resolutions, the temporal error became dominant, which showed to be second order of convergence.
    
\begin{table}
\caption{Numerical errors of the smooth density propagation benchmark.} \label{tab:sdp}
\begin{tabular}{m{2.1cm} m{3cm} l}
Resolution       & error $\lVert \rho - \rho_{ref} \rVert_{\infty}$ & order \\ \hline
$16 \times 16$   &  2.10599e-2              &       \\
$32 \times 32$   &  1.32533e-3              &  4.0  \\
$64 \times 64$   &  5.36948e-5              &  4.6  \\
$128 \times 128$ &  4.68312e-6              &  3.5  \\
$256 \times 256$ &  7.90318e-7             &  2.6 \\
$512 \times 512$ &  2.14017e-7                & 1.9 \\\hline
\end{tabular}
\end{table}

\subsection{Sod shock tube} \label{sec:shocktube}

In a first test case with discontinuities, the well-known Sod shock tube \cite{Sod1978} was simulated. The one-dimensional domain $x\in[0,1]$ was divided into two regions at $x=0.5$ with the initial conditions

\begin{eqnarray*}
    \rho_0 = 8, u_0=0, P_0 = 10, \\
    \rho_1 = 1,u_1 = 0, P_1 = 1,
\end{eqnarray*}

with $P=\rho R T$ being the pressure according to the ideal gas law and $R=1$. The periodic domain was discretized using 800 grid points; the number of cells was 200 for $p=4.$ The kinematic viscosity, Prandtl number and adiabatic exponent were set to $\nu=0.00012$, $Pr=1.0$ and $\gamma=1.4,$ respectively. With a time step size of $\delta_t=0.002$ this resulted in a relaxation time of $\tau = \nu_\mathrm{LB} / (c_s^2\delta_t) + 0.5 = 1.518.$ 


The results accurately matched the reference solution \cite{Sod1978} with only minor oscillations visible, comparable to other recently introduced methods \cite{Saadat2020}.  

\begin{figure}
    \centering
    \includegraphics[width=\linewidth]{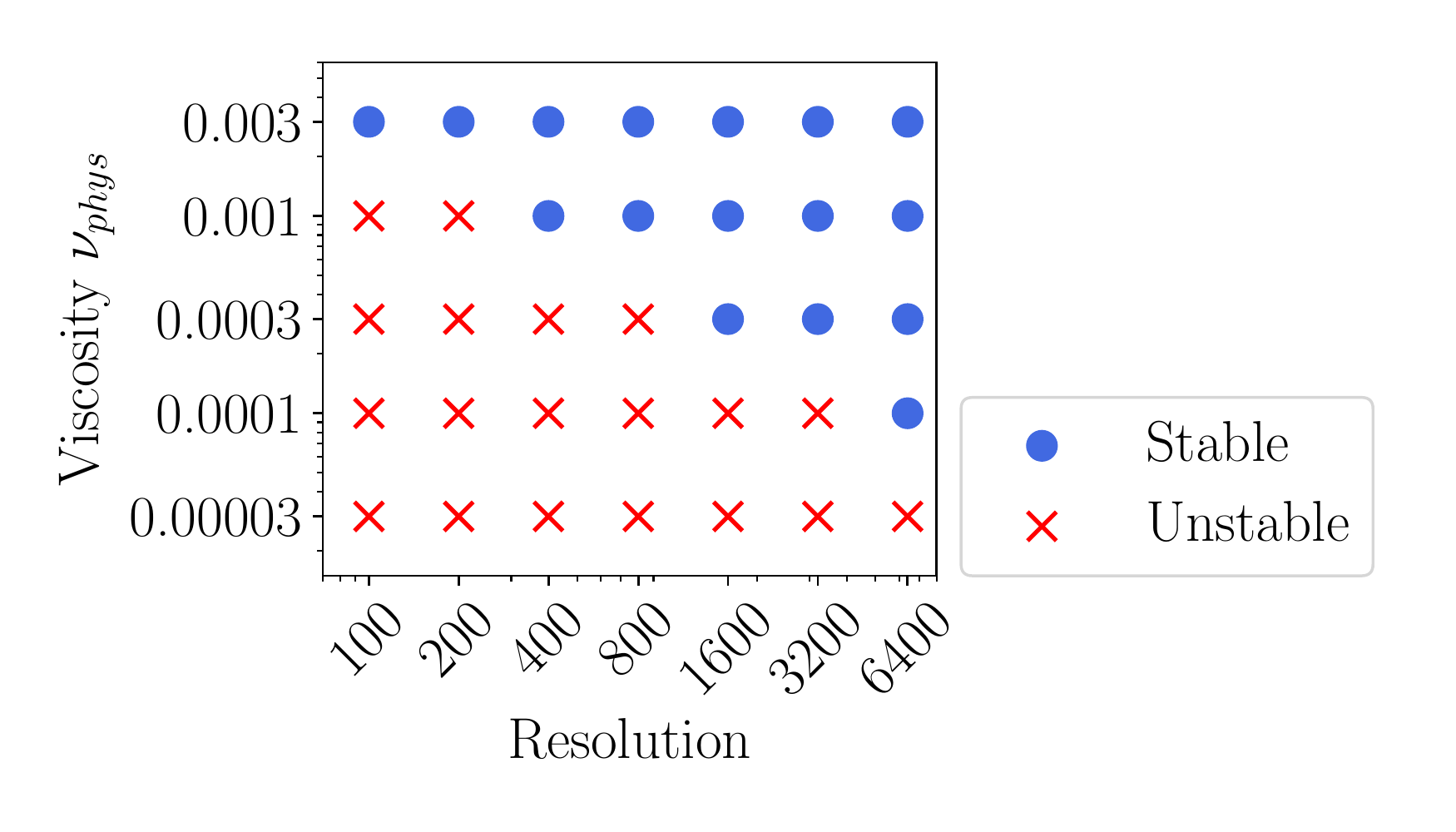}
    \caption{Stable and unstable Sod shock tube simulations of the on-lattice Boltzmann solver detailed in Subsection \ref{sec:on-lattice}.}
    \label{On-lattice-stability}
\end{figure}
\begin{figure}
    \centering
    \includegraphics[width=\linewidth]{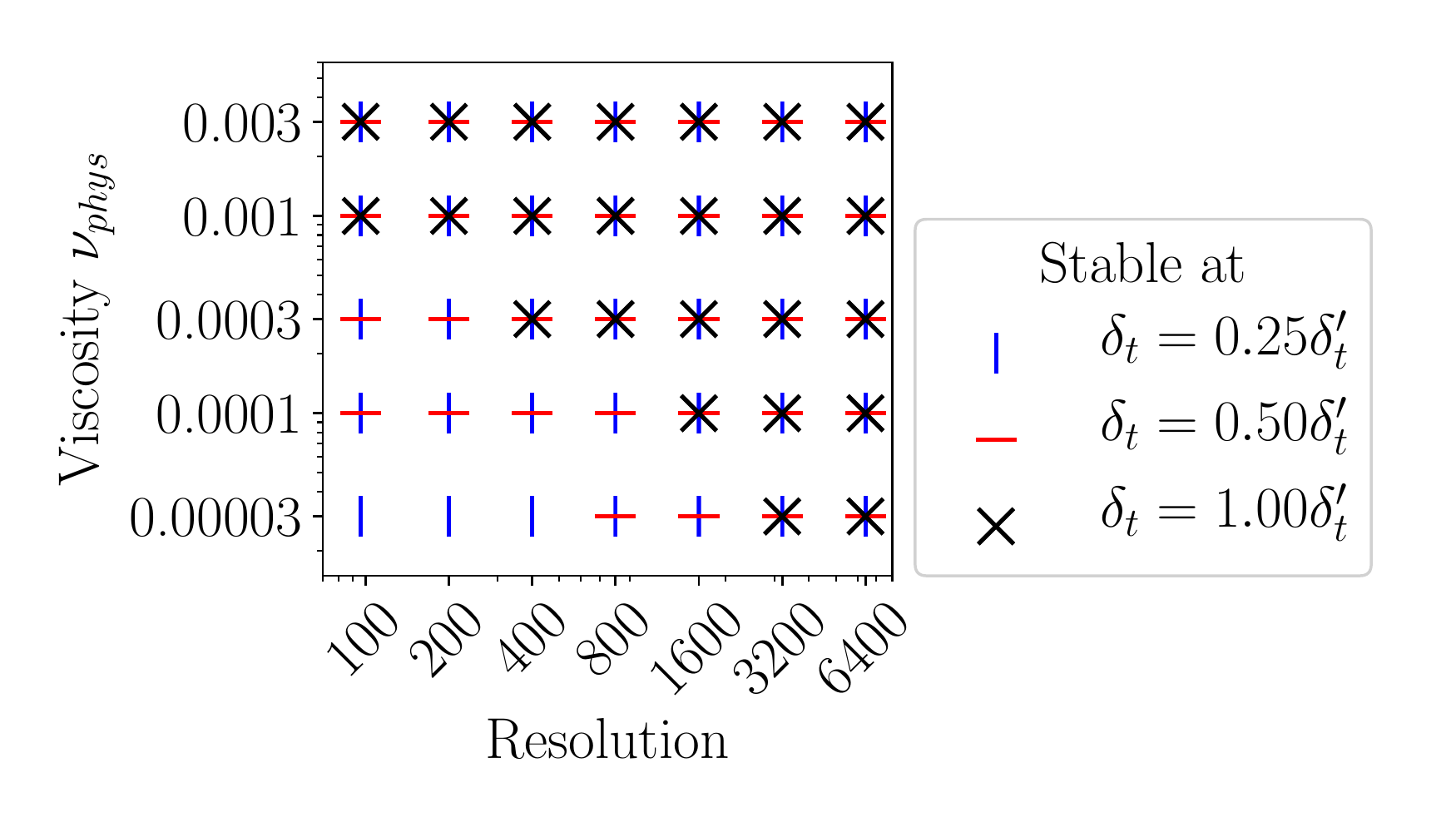}
    \caption{Stable Sod shock tube simulations of the SLLBM at three different time step sizes, where $\delta_t'$ is the time step size of the on-lattice method in Fig. \ref{On-lattice-stability}.}
    \label{Off-lattice-stability}
\end{figure}

The proposed method in our tests demonstrated clearly improved numerical stability 
compared to previous on-lattice methods. To compare the stability properties of the SLLBM and of the representative on-lattice Boltzmann solver, specified in subsection \ref{sec:on-lattice}, the Sod shock tube was again simulated for a variety of resolutions and physical viscosities, $\nu_{phys}$, until $t=1.5$. 
A simulation was rated as stable, if the simulation finished successfully and all macroscopic values were still bounded. Fig. \ref{On-lattice-stability} shows that the on-lattice Boltzmann solver was unstable for most of the lower viscosities at low resolutions. A simulation at $\nu=0.0001$, an example of an often used setting for Sod shock tubes in the literature, is only stable for the on-lattice LBM using an excessive resolution of $N=6400$.
Contrary to that, the flexible time step size of the SLLBM allowed for stable simulations even at low resolutions. Fig. \ref{Off-lattice-stability} depicts all stable simulations for three different time step sizes $\delta_t=0.25\delta_t'$, $\delta_t=0.50\delta_t'$, and $\delta_t=1.00\delta_t'$, where $\delta_t'$ is the time step size of the respective on-lattice simulation. All of the representative resolution and viscosity scenarios could be simulated stably with at least one time step size by the SLLBM.

Taken together, the Sod shock tube shows that the proposed SLLBM solver is capable to accurately capture shocks and convinces by a larger stability region than the D2V37 on-lattice Boltzmann simulation.


\subsection{2D Riemann} \label{sec:Riemann}
The two-dimensional Riemann problem is another demanding test case for compressible inviscid flow solvers, which was deeply investigated for a variety of initial configurations by Lax and Liu \cite{Lax1998} and Kurganov and Tadmor \cite{Kurganov2002}. The original works made use of Riemann solvers and semi-discrete finite difference schemes, respectively, to obtain the reference solutions. 
\begin{table}
    \centering
    \caption{Initial conditions of the 2D Riemann problem according to Lax and Liu \cite{Lax1998}.}
    \label{tab:riemann}
\begin{tabular}{|l|l|}

\hline
{$\!\begin{aligned} 
                    
     \rho &= 1  \\
     u_x &= 0.7276 \\
     u_y &= 0  \\
     P &= 1
     \end{aligned}$} &
     {$\!\begin{aligned} 
    \rho &= 0.5313 \\
      u_x &= 0 \\
      u_y &= 0 \\
      P&=0.4
     \end{aligned}$} \\
     \hline
     {$\!\begin{aligned}
     \rho &= 0.8\\
     u_x &= 0 \\
     u_y &= 0 \\
     P &= 1
     \end{aligned}$} & 
     {$\!\begin{aligned}
      \rho &= 1 \\
      u_x &= 0 \\
      u_y &= 0.7276 \\
      P&=1
     \end{aligned}$} \\
     \hline
\end{tabular}
\end{table}



\begin{figure}
    \centering
    \includegraphics[width=0.95\linewidth]{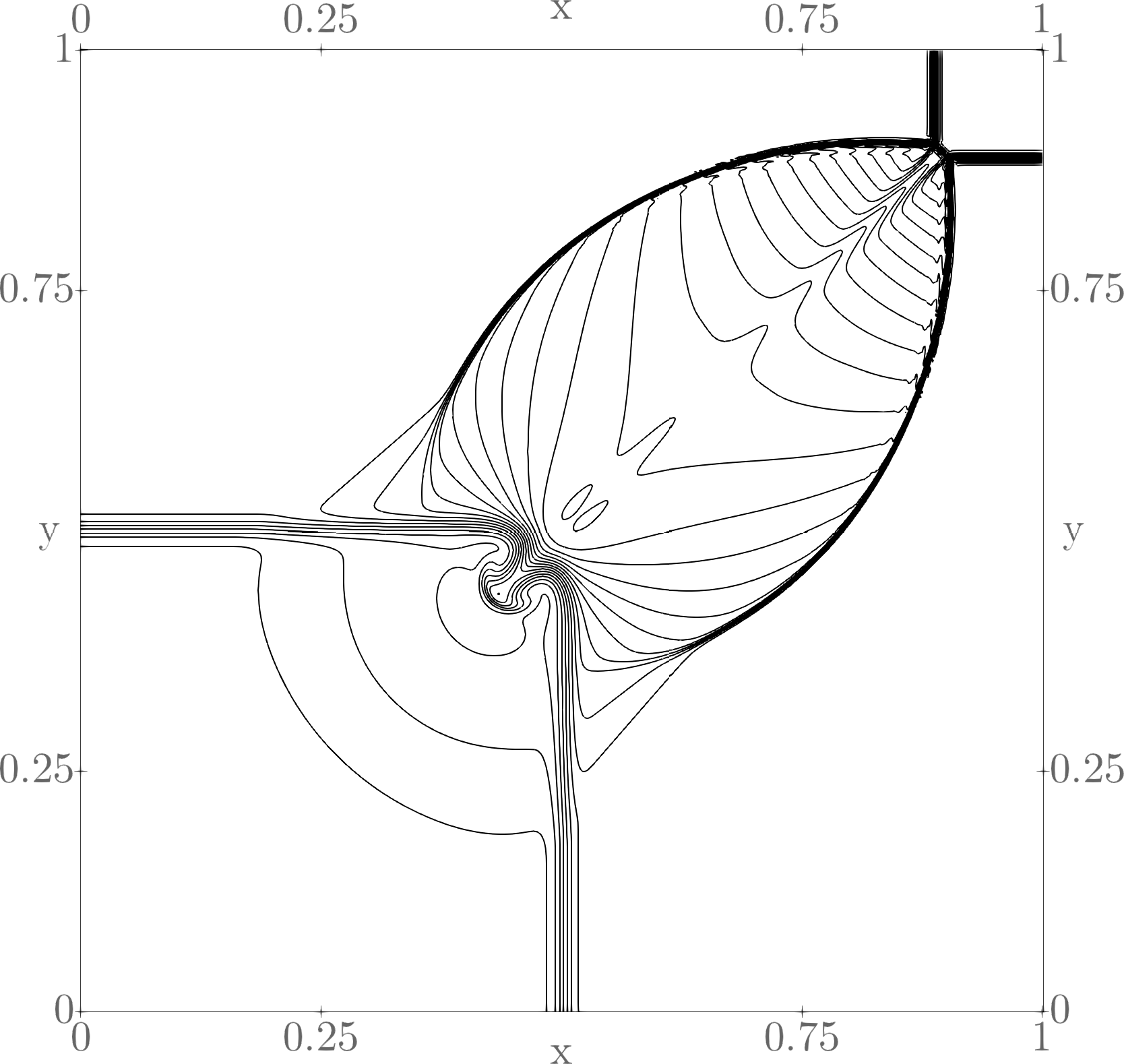}
    \caption{Outline of 45 equally spaced density contours of the 2D Riemann problem with $N=512\times512$ in the interval $\rho \in [0.412, 1.753]$. The order of finite elements was $p=4$. The simulation end time was reached in 2091 iterations. Reference solutions obtained by two different solvers are available in \cite{Lax1998} and \cite{Kurganov2002}}
    \label{fig:Riemann_glc}
\end{figure}

The domain $(x,y) \in [0,1]$ was divided into four quadrants with the initial conditions shown in Table \ref{tab:riemann}. The simulation domain was discretized into $N_\Xi = 128\times 128$ cells with an order of finite elements of $p=4$, which results in $N= 512\times512$ grid points. The relaxation time $\tau = 1.94815$ was determined according to $\tau = \nu / (c_s^2 \delta_t) + 0.5$. The physical simulation end time was $t_\mathrm{end}=0.25$, which was reached after 2091 iterations, i.e. the viscosity is $\nu=0.000173$, assuming the reference temperature $T_0 = 1$. The heat capacity ratio was set to $\gamma=1.4$.  At the boundaries, a zero-gradient condition, i.e. $\partial f = 0$, was applied. From a practical point-of-view, to free the simulation domain from disturbing boundary interactions, the domain was set to twice the size and clipped during post-processing. At first, Gauss-Lobatto-Chebyshev nodes were used to distribute the support points within each cell. Fig. \ref{fig:Riemann_glc} shows the density contours after 2091 iterations. In total, 45 equally distributed contours were applied. The results were in good agreement with the reference solution in \cite{Lax1998, Kurganov2002}, although minor oscillations were still visible near the shock fronts. 

{}

\begin{figure*}
    \centering
    \includegraphics[width=6in]{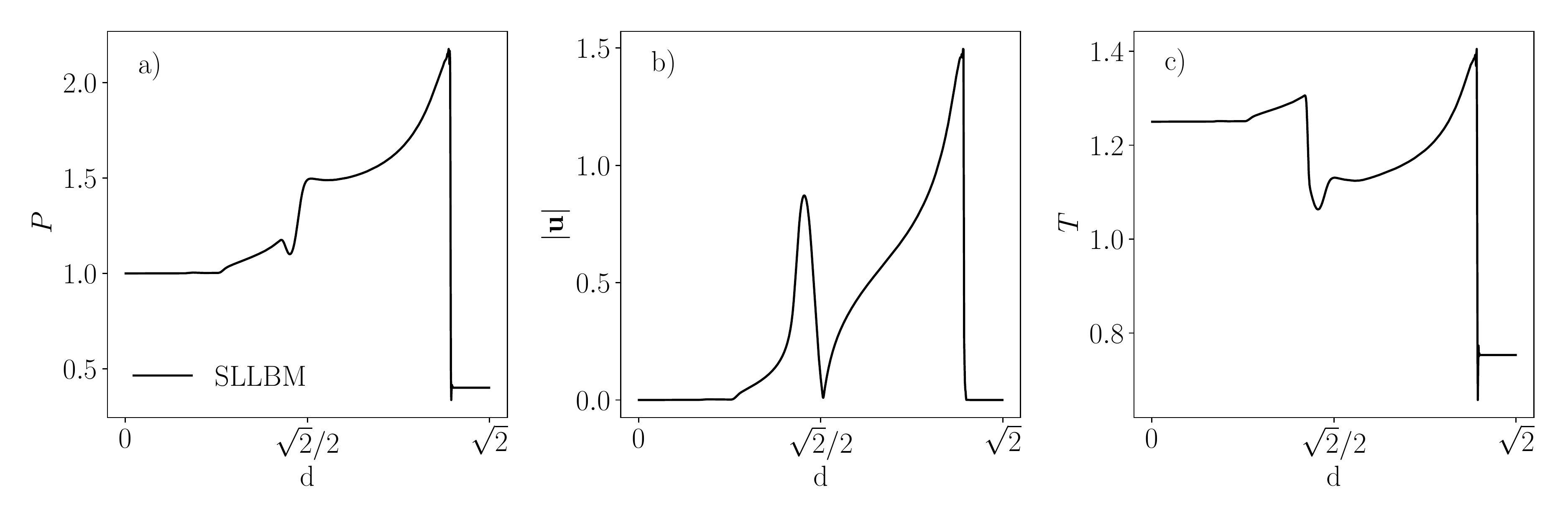}
    \caption{Pressure, velocity and temperature curves of the two-dimensional Riemann problem, shown in Fig. \ref{fig:Riemann_glc} along the $y=x$ diagonal $d$. }
    \label{fig:riemann_values}
\end{figure*}

Fig. \ref{fig:riemann_values} depicts the pressure, velocity and temperature charts along the $y=x$ diagonal at the simulation end time. Although there is no reference solution available, the plot gives a good impression of the strong discontinuities that are also the main source of instabilities in this test case. 

\begin{figure}
    \centering
    \includegraphics[width=0.8\linewidth]{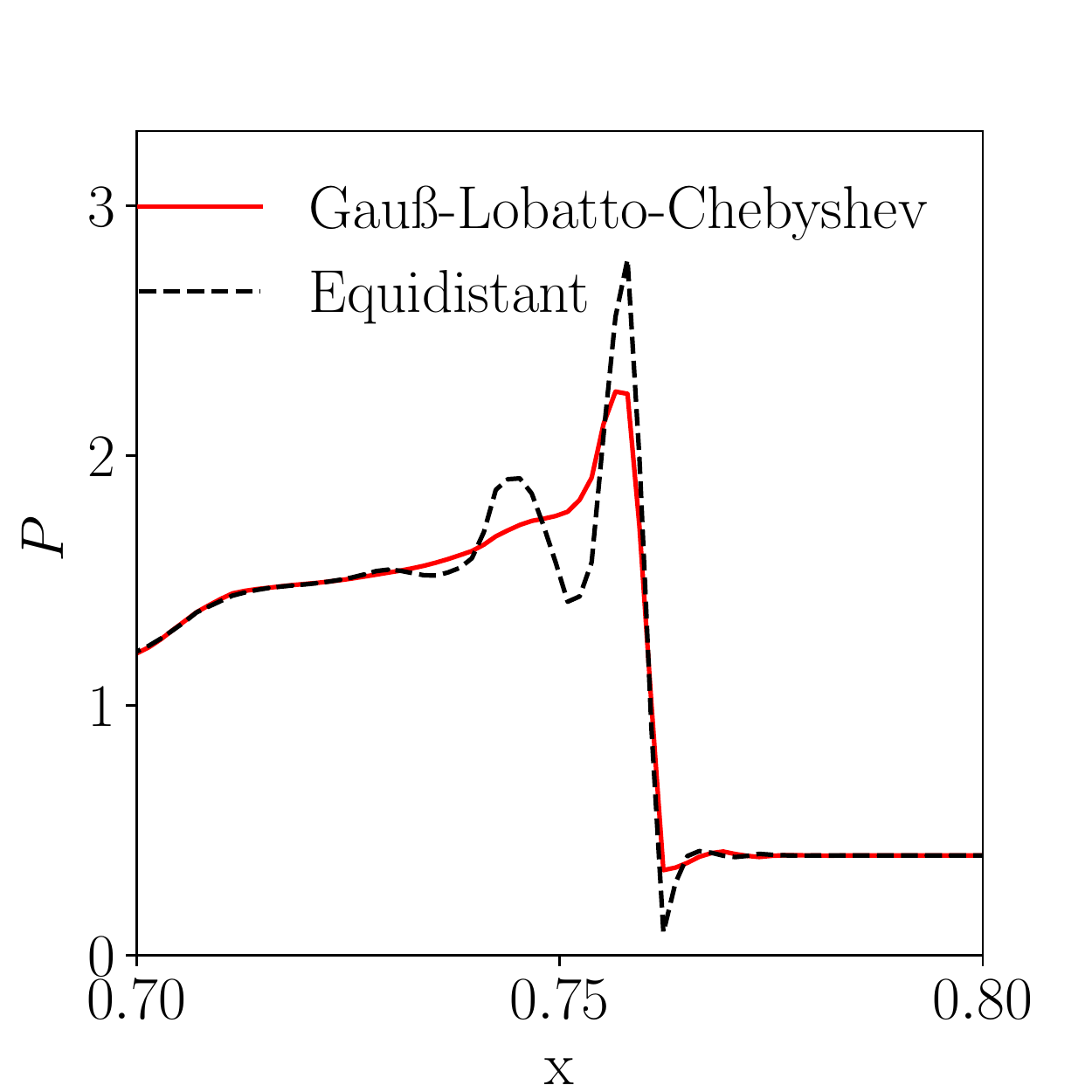}
    \caption{Comparison of the pressure $P$ with Gauß-Lobatto-Chebyshev nodes and conventional equidistant nodes at order of finite elements of $p=4$ after 200 time steps. The test run with equidistant nodes provoked heavy oscillations, which resulted in a crash shortly afterwards.}
    \label{fig:GLC_vs_Equi}
\end{figure}

To emphasize the value of non-equidistant support points for the interpolation procedure, the same configuration was simulated again with equidistant nodes. The simulation became unstable shortly after 200 time steps. Fig. \ref{fig:GLC_vs_Equi} shows the pressure along the $y=x$ axis and reveals the heavy oscillations of the simulation with equidistant nodes shortly before the simulation diverges.

{}

\subsection{\label{sec:svi} Shock vortex interaction}

The final test case was a 2D shock vortex interaction, which was originally presented by Inoue and Hattori \cite{Inoue1999}. For this simulation a transformed grid was used, which shows the applicability of the SLLBM to non-uniform grids. The simulation set-up was as follows: the domain consists of two regions, which are seperated by a steady shock. The Mach number of region A is $Ma_A=-1.2$, while the Mach number in region B $Ma_B\approx-0.842$ is determined by the Rankine-Hugoniot condition. A vortex with characteristic Mach number $Ma_v=0.5$ is advected in region A towards the shock (from right to left) and interacts with it throughout the simulation.
The flow field of the vortex is given by:
\begin{equation}
    u_\theta(r)=\sqrt{\gamma T}\mathrm{Ma_v} r \mathrm{exp}((1-r^2)/2)
\end{equation}
The initial pressure and density field are 
\begin{equation}
    P(r)=\frac{1}{\gamma}\left(1-\frac{\gamma-1}{2}\mathrm{Ma_v}\mathrm{exp}(1-r^2)\right)^{\gamma/(\gamma-1)}
\end{equation}
and 
\begin{equation}
    \rho(r)=\left(1-\frac{\gamma-1}{2}\mathrm{Ma_v}\mathrm{exp}(1-r^2)\right)^{1/(\gamma-1)}.
\end{equation}

The temperature field can be determined via the ideal gas law $P = \rho R T$.

The Reynolds number is defined by  
\begin{equation}
Re=400=\frac{c_s R}{ \nu}
\end{equation}
with $R$ being the resolution of the characteristic radius $r$. The Prandtl number is $Pr=0.75$.

\begin{figure}
    \centering
\includegraphics[width=0.9\linewidth]{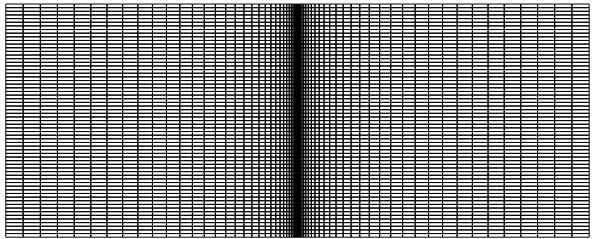}
    \caption{Schematic grid for the shock vortex interaction. The actual grid used in the simulation was further refined by splitting each cell in half along each axis. Each cell contains $5\times5$ support points, whereby the outer support points are shared with the neighboring cells.}
    \label{fig:grid}
\end{figure}

The simulation was run on a physical domain of size $60\times24$, which was discretized by $N_\Xi=256\times256$ cells and the polynomial order was $p=4$, resulting in 1024 lattice nodes in each direction. However, to highly resolve the steady shock the grid was transformed in x-direction according to a simple transformation function:
\begin{equation}
 x_\mathrm{trans}=\frac{30}{\pi}\left({\Lambda \sin\left(\frac{\pi}{30} (x)\right) + \frac{\pi}{30}x}\right),
\end{equation}

where $\Lambda$ is a stretching parameter, which was set to $\Lambda=0.95$. A similar refinement was done by Inoue and Hattori in the original work \cite{Inoue1999}. Although the shock slightly moved during the interaction with the shock, the shock was still located in the refinement zone, which led to satisfactory results and reduced the amount of needed grid points. The shock was located at $x_\mathrm{shock}=30$ and the initial vortex center position was $x_\mathrm{vortex}=32$. We observed that the DNS results by Inoue and Hattori were only matched if the vortex was initialized exactly at this position although partly located in the shock. The schematic grid of the computation is shown in Fig. \ref{fig:grid}. The average resolution of the characteristic radius along the x-axis $r$ was $1024/60\approx17.07$ grid points and $1024/24\approx42.67$ grid points along the y-axis. The large domain and periodic boundary conditions in all directions were sufficient to keep the domain of interest free from disturbing influences of the boundaries. The characteristic time is defined as $t'={R}/{c_s}$. The time step size was $\delta_t = 7 \cdot 10^{-4}$ so that the final result at $t=8$ was reached in 11290 iterations.

\begin{figure}
    \centering
\includegraphics[width=0.95\linewidth]{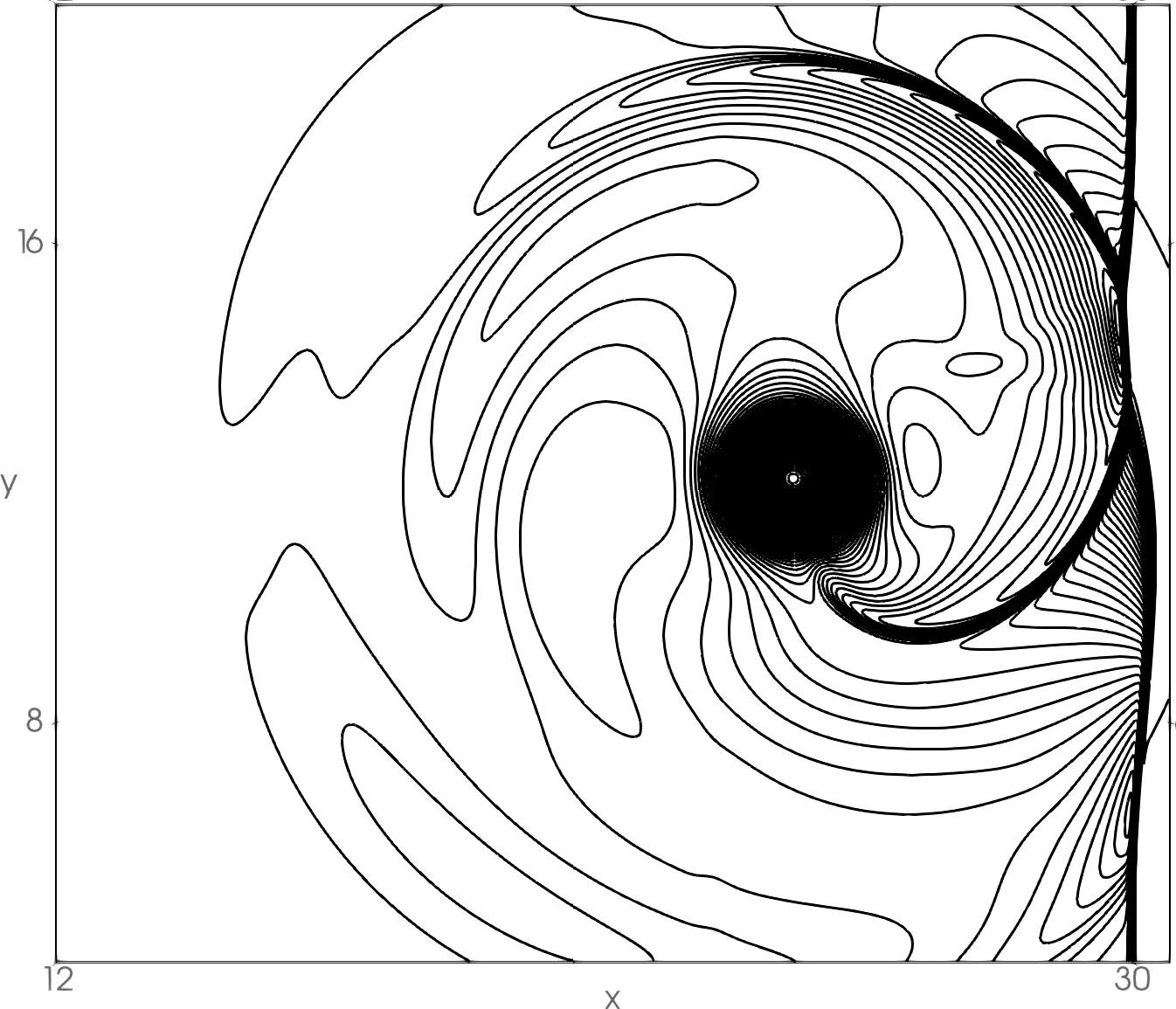}
    \caption{Density contours of the shock-vortex interaction with $Ma_v = 0.5$ and $Re=400$ in the range $\rho \in [0.92,1.55]$ in 119 steps.}
    \label{fig:svi}
\end{figure}

Fig. \ref{fig:svi} shows the density contours in the range $\rho \in [0.92, 1.55]$ with 119 contours. The contours were in good agreement with previous works, e.g. \cite{Inoue1999,Frapolli2017}. Both the static shock and the attached shock were well resolved and the density contours appeared without any oscillations.

\begin{figure}[htp]
    \centering
\includegraphics[width=\linewidth]{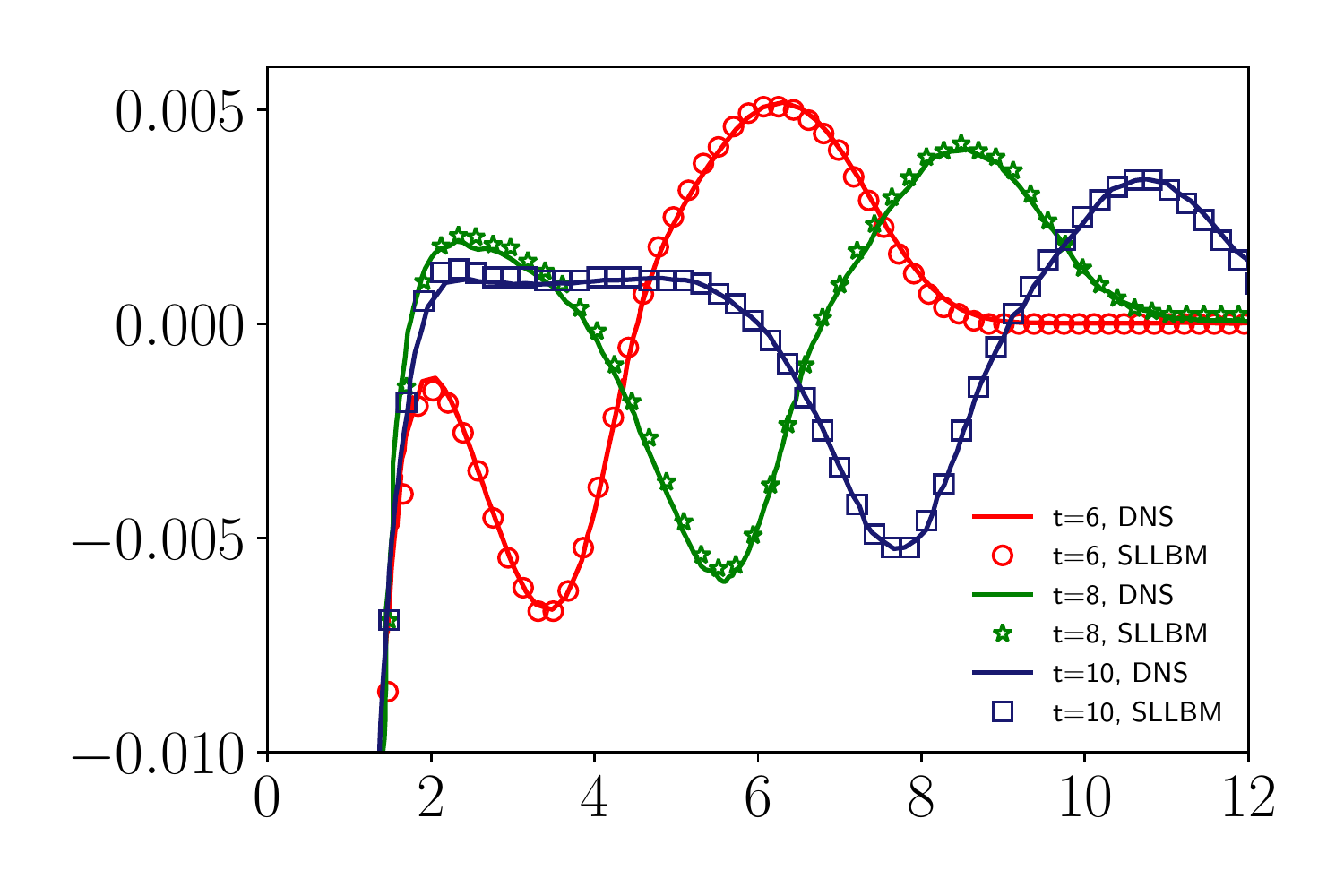}
    \caption{Sound pressure $\Delta P = (P - P_B) / P_B$ of the shock-vortex interaction with $Re=800, Ma_v=0.25$ and $Ma_A=1.2$ at times $t'=6,$ $t'=8,$ and $t'=10$. The measurements were taken from the center of the vortex along the radius with $-45^{\circ}$ with respect to the x-axis. DNS from \cite{Inoue1999}.}
    \label{fig:radial}
\end{figure}

To quantify the results, the radial sound pressure in a simulation with $Ma_v=0.25$ and $Re=800$ was evaluated and compared with DNS results from Inoue and Hattori \cite{Inoue1999}. Fig. \ref{fig:radial} shows the normalized pressure $\Delta P = (P - P_B) / P_B)$, where  $P_B$ denotes the initial pressure in region B. The results of the present configuration were in excellent agreement with the reference solution obtained by the DNS.

\begin{figure}
    \centering
    \includegraphics[width=0.8\linewidth]{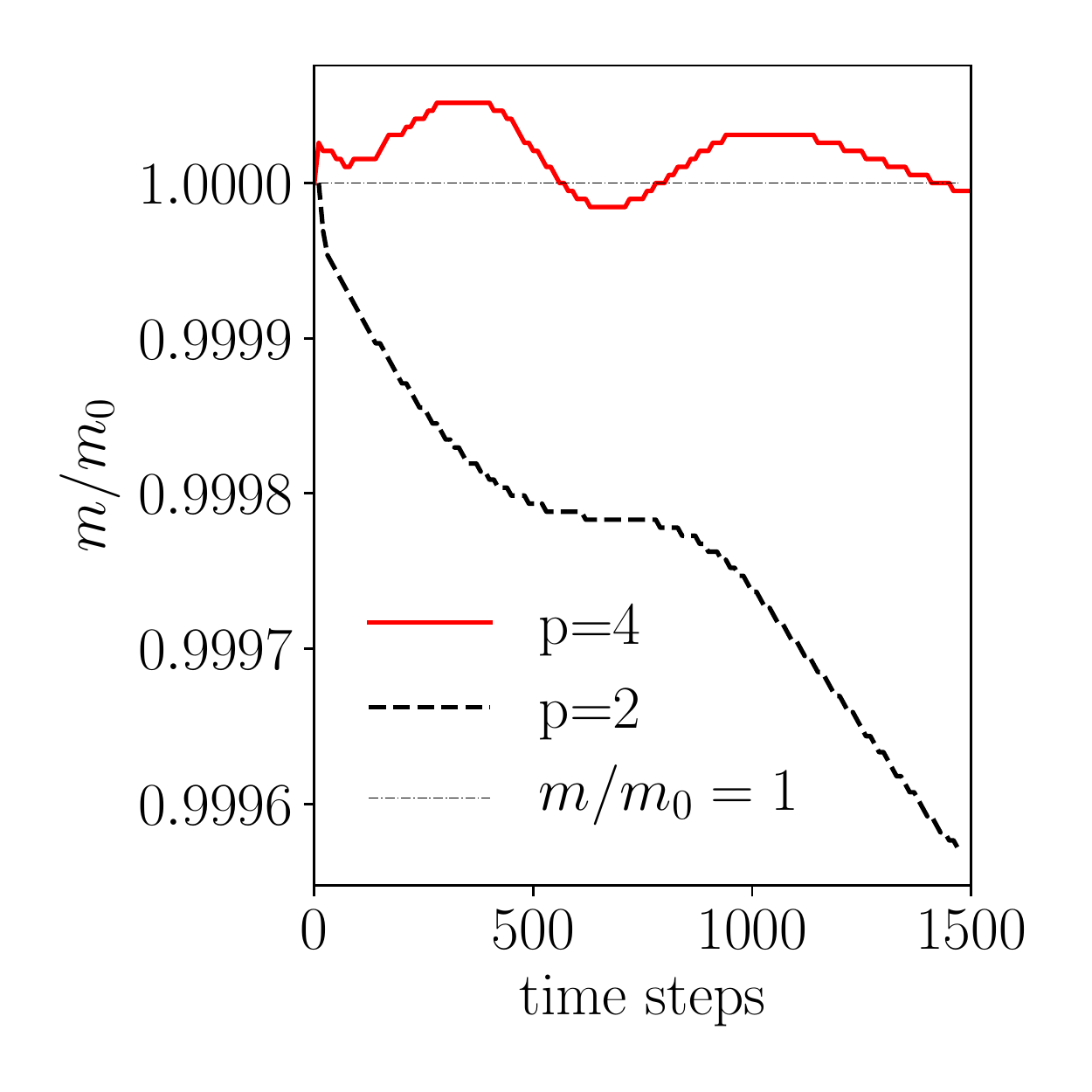}
    \caption{Relative mass $m/m_0$ for polynomial orders $p=4$ and $p=2$ over the number of iterations for the test case shown in Fig. \ref{fig:svi}.}
    \label{fig:mass_conservation}
\end{figure}

The total mass in the previously presented well-resolved test cases was found to be more or less constant over the entire simulation time. However, since the SLLBM is not rigorously mass-conserving, Fig. \ref{fig:mass_conservation} presents the evolution of mass over 1500 simulation steps for a slightly underresolved configuration of $Re=800$, $Ma=0.25$ and $512 \times 512$ grid points at two polynomial orders. In case of a low-order polynomial interpolation of $p=2$, there was a minor but steady loss of mass, which amounted to approximately 0.05\% of the initial mass after 1500 steps. The mass conservation was tighter for the high-order interpolation. For $p=4,$ the mass oscillated around the initial value as opposed to a systematic gain or loss.

\section{Discussion}
\label{sec:discussion}

As detailed in the introduction of this work, the research on lattice Boltmann methods for compressible flows is ongoing. We have shown how the use of an efficient off-lattice streaming step enables compressible simulations by relaxing the coupling between time step, computational grid, and particle velocities in the LBM.

Originating from the usual lattice Boltzmann equation the SLLBM algorithm follows the trajectories along the characteristics and recovers the off-lattice departure points using a finite element interpolation. In common with finite volume or finite difference LBMs, this procedure allows for a flexible time step size. This key feature is missing in usual on-lattice Boltzmann solvers for compressible flows. Contrary to the Eulerian methods, the SLLBM avoids costly and often high-dissipative time integration. The compressible SLLBM is related to the PoD method, but operates in a static, non-moving reference frame. Despite this regression, the SLLBM was capable to simulate a variety of compressible test cases. As the weights in the present model are not temperature-dependent, the temperature ratio is - in principle - not limited. Simple temperature diffusion tests not shown in this article revealed a stable temperature ratio of at least 1:1000.

We tested the compressible SLLBM for five common test cases, in both inviscid and viscous regimes. The two dimensional Taylor-Green vortex showed the high degree of Galilean invariance of the scheme. The accuracy test by a smooth density propagation confirmed the fourth order in space and second order in time of the scheme. The Sod shock tube in its original configuration and a pressure ratio of $P_1/P_2=10$  was successfully tested with the SLLBM. In contrast, on-lattice Boltzmann solvers like the one presented in \cite{Saadat2019} often restrict the pressure ratio to lower values on larger grids, while a recursive regularization allowed to assure stability in the work by Coreixas et al. \cite{Coreixas2017}. In contrast, the SLLBM's variable time step size allows to adapt the time step to the needs of the simulation, even with the most common BGK model.
Furthermore, we employed fourth order interpolation polynomials, rendering the solution in the smooth areas spatially to the corresponding order \cite{Kramer2017a}, which is another unique property of this scheme. However, as shown in the two-dimensional Riemann problem in section \ref{sec:Riemann}, leveraging the spatial order is only workable for an appropriate distribution of support points within each cell. While equidistant support points led to unstable simulations, the use of fourth order Gauss-Lobatto-Chebyshev points enabled stable simulations.  Moreover, the simulation of the shock-vortex interaction showed that a high-order interpolation improves mass conservation, which is not rigorously enforced in interpolation-based LBM \cite{Zarikov2019}.
The equilibrium distribution function based on a Hermite expansion of fourth order is computationally intense, but allows to represent compressible and viscous flows with energy conservation \cite{Shan2006}. Concerning the discretization of the velocity space, a D2Q25 velocity set based on the fifth-order Hermite polynomial has been used, which is suited to recover the moments up to fourth order correctly \cite{Chikatamarla2009}. In contrast, on-lattice velocity sets usually provoke deviations in the recovery of the high-order moments \cite{Chikatamarla2009}. The combination of the equilibrium and velocity set allowed for a total Mach number of $\mathrm{Ma}=1.7$ in the simulation of the shock-vortex interaction, but the results of the 2D Taylor-Green vortex flow suggest even higher Mach numbers, depending on the time step size. A stability analysis \cite{Sterling1996, Hosseini2019a} could clarify the exact limits of the scheme. An obvious way to further increase the Mach number is to incorporate statically shifted \cite{Frapolli2016, Hosseini2019} or dynamically shifted lattices \cite{Dorschner2018}. 
Future works should thoroughly investigate the path to extend the velocity sets to higher velocities in three dimensions, while keeping the number of discrete velocities lower than 125. The effects of lattice pruning, which were thoroughly studied in the weakly compressible case, e.g. D3Q27 to D3Q19 \cite{White2011}, should also be examined for these high-order velocity sets. Besides this, the extensions of co-moving reference frames accompanied by high-order polynomial orders should also be evaluated \cite{Dorschner2018}. 
Additionally, we propose to pursue an approach that was originally supposed by Watari and Hattori \cite{Watari2006} for finite difference LBMs. The authors combined platonic solids, dodecahedra and icosahedra, to construct highly isotropic velocity sets at different velocity shells. It seems to be worth investigating if these velocity also allow us to increase the Mach number in SLLBM simulations. Corresponding results will be published in an upcoming article.

Another issue is the second distribution function value, needed to correctly represent the heat capacity ratio $\gamma$. This model was also successfully used by other authors \cite{Frapolli2015,Frapolli2017,Saadat2019}. A reduction of the velocity set size of $g$ could possibly make the compressible LBM more attractive.

Taken together, the presented SLLBM opens an alternative and attractive field of solvers for compressible flows and, at the same time, combines many insights from previous LBM solvers for compressible flows.

\section{Conclusions}
\label{sec:conclusion}
In this article we showed that the semi-Lagrangian streaming step enables efficient lattice Boltzmann simulations of compressible flows even in a static, non-moving reference frame. In contrast to other methods, the vertices are organized in cells, and interpolation polynomials up to fourth order are used to attain the off-vertex distribution function values.

The presented SLLBM solver for compressible flows convinces by six main advantages:
\begin{enumerate}
    \item flexibility of the time step size,
    \item absence of a dedicated time integrator as the integration is performed along characteristics,
    \item applicability to stretched or unstructured meshes
    \item possibility to employ advantageous velocity sets
    \item limitation of mass loss by high-order polynomials, and
    \item fourth order of convergence in space and second order in time.
\end{enumerate}

The test cases of a 2D Taylor Green vortex, a smooth density propagation, a Sod shock tube, a 2D Riemann problem and a shock-vortex interaction on transformed grids show the general applicability to flows with discontinuities. Thereby, Gauss-Lobatto-Chebyshev support points allow to reduce oscillations near the shock fronts.

In our future work, the presented framework will applied to three-dimensional test cases with a focus on further increasing the time step size and on reducing the number of discrete velocities. 

\acknowledgments
The authors would like to thank Ilya V. Karlin, Seyed Ali Hosseini, and Christophe Coreixas for fruitful discussions and for proofreading the manuscript.
This work was supported by the German Ministry of Education and Research and the Ministry for Culture and Science North Rhine-Westfalia (research grant 13FH156IN6). D. Wilde is supported by German Research Foundation (DFG) project FO 674/17-1.

\bibliography{compressible}

\begin{thebibliography}{68}%
\makeatletter
\providecommand \@ifxundefined [1]{%
 \@ifx{#1\undefined}
}%
\providecommand \@ifnum [1]{%
 \ifnum #1\expandafter \@firstoftwo
 \else \expandafter \@secondoftwo
 \fi
}%
\providecommand \@ifx [1]{%
 \ifx #1\expandafter \@firstoftwo
 \else \expandafter \@secondoftwo
 \fi
}%
\providecommand \natexlab [1]{#1}%
\providecommand \enquote  [1]{``#1''}%
\providecommand \bibnamefont  [1]{#1}%
\providecommand \bibfnamefont [1]{#1}%
\providecommand \citenamefont [1]{#1}%
\providecommand \href@noop [0]{\@secondoftwo}%
\providecommand \href [0]{\begingroup \@sanitize@url \@href}%
\providecommand \@href[1]{\@@startlink{#1}\@@href}%
\providecommand \@@href[1]{\endgroup#1\@@endlink}%
\providecommand \@sanitize@url [0]{\catcode `\\12\catcode `\$12\catcode
  `\&12\catcode `\#12\catcode `\^12\catcode `\_12\catcode `\%12\relax}%
\providecommand \@@startlink[1]{}%
\providecommand \@@endlink[0]{}%
\providecommand \url  [0]{\begingroup\@sanitize@url \@url }%
\providecommand \@url [1]{\endgroup\@href {#1}{\urlprefix }}%
\providecommand \urlprefix  [0]{URL }%
\providecommand \Eprint [0]{\href }%
\providecommand \doibase [0]{http://dx.doi.org/}%
\providecommand \selectlanguage [0]{\@gobble}%
\providecommand \bibinfo  [0]{\@secondoftwo}%
\providecommand \bibfield  [0]{\@secondoftwo}%
\providecommand \translation [1]{[#1]}%
\providecommand \BibitemOpen [0]{}%
\providecommand \bibitemStop [0]{}%
\providecommand \bibitemNoStop [0]{.\EOS\space}%
\providecommand \EOS [0]{\spacefactor3000\relax}%
\providecommand \BibitemShut  [1]{\csname bibitem#1\endcsname}%
\let\auto@bib@innerbib\@empty
\bibitem [{\citenamefont {Dorschner}\ \emph {et~al.}(2018)\citenamefont
  {Dorschner}, \citenamefont {B\"osch},\ and\ \citenamefont
  {Karlin}}]{Dorschner2018}%
  \BibitemOpen
  \bibfield  {author} {\bibinfo {author} {\bibfnamefont {B.}~\bibnamefont
  {Dorschner}}, \bibinfo {author} {\bibfnamefont {F.}~\bibnamefont {B\"osch}},
  \ and\ \bibinfo {author} {\bibfnamefont {I.~V.}\ \bibnamefont {Karlin}},\
  }\href {\doibase 10.1103/PhysRevLett.121.130602} {\bibfield  {journal}
  {\bibinfo  {journal} {Phys. Rev. Lett.}\ }\textbf {\bibinfo {volume} {121}},\
  \bibinfo {pages} {130602} (\bibinfo {year} {2018})}\BibitemShut {NoStop}%
\bibitem [{\citenamefont {McNamara}\ and\ \citenamefont
  {Zanetti}(1988)}]{McNamara1988}%
  \BibitemOpen
  \bibfield  {author} {\bibinfo {author} {\bibfnamefont {G.~R.}\ \bibnamefont
  {McNamara}}\ and\ \bibinfo {author} {\bibfnamefont {G.}~\bibnamefont
  {Zanetti}},\ }\href {\doibase 10.1103/PhysRevLett.61.2332} {\bibfield
  {journal} {\bibinfo  {journal} {Phys. Rev. Lett.}\ } (\bibinfo {year}
  {1988}),\ 10.1103/PhysRevLett.61.2332}\BibitemShut {NoStop}%
\bibitem [{\citenamefont {Higuera}\ and\ \citenamefont
  {Jim{\'{e}}nez}(1989)}]{Higuera1989}%
  \BibitemOpen
  \bibfield  {author} {\bibinfo {author} {\bibfnamefont {F.~J.}\ \bibnamefont
  {Higuera}}\ and\ \bibinfo {author} {\bibfnamefont {J.}~\bibnamefont
  {Jim{\'{e}}nez}},\ }\href {\doibase 10.1209/0295-5075/9/7/009} {\bibfield
  {journal} {\bibinfo  {journal} {EPL}\ } (\bibinfo {year} {1989}),\
  10.1209/0295-5075/9/7/009}\BibitemShut {NoStop}%
\bibitem [{\citenamefont {Chen}\ \emph {et~al.}(1991)\citenamefont {Chen},
  \citenamefont {Chen}, \citenamefont {Martnez},\ and\ \citenamefont
  {Matthaeus}}]{Chen1991}%
  \BibitemOpen
  \bibfield  {author} {\bibinfo {author} {\bibfnamefont {S.}~\bibnamefont
  {Chen}}, \bibinfo {author} {\bibfnamefont {H.}~\bibnamefont {Chen}}, \bibinfo
  {author} {\bibfnamefont {D.}~\bibnamefont {Martnez}}, \ and\ \bibinfo
  {author} {\bibfnamefont {W.}~\bibnamefont {Matthaeus}},\ }\href {\doibase
  10.1103/PhysRevLett.67.3776} {\bibfield  {journal} {\bibinfo  {journal}
  {Phys. Rev. Lett.}\ } (\bibinfo {year} {1991}),\
  10.1103/PhysRevLett.67.3776}\BibitemShut {NoStop}%
\bibitem [{\citenamefont {Alexander}\ \emph {et~al.}(1992)\citenamefont
  {Alexander}, \citenamefont {Chen}, \citenamefont {Chen},\ and\ \citenamefont
  {Doolen}}]{Alexander1992}%
  \BibitemOpen
  \bibfield  {author} {\bibinfo {author} {\bibfnamefont {F.~J.}\ \bibnamefont
  {Alexander}}, \bibinfo {author} {\bibfnamefont {H.}~\bibnamefont {Chen}},
  \bibinfo {author} {\bibfnamefont {S.}~\bibnamefont {Chen}}, \ and\ \bibinfo
  {author} {\bibfnamefont {G.~D.}\ \bibnamefont {Doolen}},\ }\href {\doibase
  10.1103/PhysRevA.46.1967} {\bibfield  {journal} {\bibinfo  {journal}
  {Physical Review A}\ }\textbf {\bibinfo {volume} {46}},\ \bibinfo {pages}
  {1967} (\bibinfo {year} {1992})}\BibitemShut {NoStop}%
\bibitem [{\citenamefont {Kr{\"{u}}ger}\ \emph {et~al.}(2017)\citenamefont
  {Kr{\"{u}}ger}, \citenamefont {Kusumaatmaja}, \citenamefont {Kuzmin},
  \citenamefont {Shardt}, \citenamefont {Silva},\ and\ \citenamefont
  {Viggen}}]{Kruger2016}%
  \BibitemOpen
  \bibfield  {author} {\bibinfo {author} {\bibfnamefont {T.}~\bibnamefont
  {Kr{\"{u}}ger}}, \bibinfo {author} {\bibfnamefont {H.}~\bibnamefont
  {Kusumaatmaja}}, \bibinfo {author} {\bibfnamefont {A.}~\bibnamefont
  {Kuzmin}}, \bibinfo {author} {\bibfnamefont {O.}~\bibnamefont {Shardt}},
  \bibinfo {author} {\bibfnamefont {G.}~\bibnamefont {Silva}}, \ and\ \bibinfo
  {author} {\bibfnamefont {E.~M.}\ \bibnamefont {Viggen}},\ }\href {\doibase
  10.1007/978-3-319-44649-3} {\emph {\bibinfo {title} {{The lattice Boltzmann
  method: Principles and practice}}}}\ (\bibinfo {year} {2017})\BibitemShut
  {NoStop}%
\bibitem [{\citenamefont {Frapolli}\ \emph {et~al.}(2015)\citenamefont
  {Frapolli}, \citenamefont {Chikatamarla},\ and\ \citenamefont
  {Karlin}}]{Frapolli2015}%
  \BibitemOpen
  \bibfield  {author} {\bibinfo {author} {\bibfnamefont {N.}~\bibnamefont
  {Frapolli}}, \bibinfo {author} {\bibfnamefont {S.~S.}\ \bibnamefont
  {Chikatamarla}}, \ and\ \bibinfo {author} {\bibfnamefont {I.~V.}\
  \bibnamefont {Karlin}},\ }\href {\doibase 10.1103/PhysRevE.92.061301}
  {\bibfield  {journal} {\bibinfo  {journal} {Phys. Rev. E}\ }\textbf {\bibinfo
  {volume} {92}} (\bibinfo {year} {2015}),\
  10.1103/PhysRevE.92.061301}\BibitemShut {NoStop}%
\bibitem [{\citenamefont {Atif}\ \emph {et~al.}(2018)\citenamefont {Atif},
  \citenamefont {Namburi},\ and\ \citenamefont {Ansumali}}]{Atif2018}%
  \BibitemOpen
  \bibfield  {author} {\bibinfo {author} {\bibfnamefont {M.}~\bibnamefont
  {Atif}}, \bibinfo {author} {\bibfnamefont {M.}~\bibnamefont {Namburi}}, \
  and\ \bibinfo {author} {\bibfnamefont {S.}~\bibnamefont {Ansumali}},\ }\href
  {\doibase 10.1103/PhysRevE.98.053311} {\bibfield  {journal} {\bibinfo
  {journal} {Phys. Rev. E}\ }\textbf {\bibinfo {volume} {98}},\ \bibinfo
  {pages} {053311} (\bibinfo {year} {2018})}\BibitemShut {NoStop}%
\bibitem [{\citenamefont {Saadat}\ \emph {et~al.}(2019)\citenamefont {Saadat},
  \citenamefont {B\"osch},\ and\ \citenamefont {Karlin}}]{Saadat2019}%
  \BibitemOpen
  \bibfield  {author} {\bibinfo {author} {\bibfnamefont {M.~H.}\ \bibnamefont
  {Saadat}}, \bibinfo {author} {\bibfnamefont {F.}~\bibnamefont {B\"osch}}, \
  and\ \bibinfo {author} {\bibfnamefont {I.~V.}\ \bibnamefont {Karlin}},\
  }\href {\doibase 10.1103/PhysRevE.99.013306} {\bibfield  {journal} {\bibinfo
  {journal} {Phys. Rev. E}\ }\textbf {\bibinfo {volume} {99}},\ \bibinfo
  {pages} {013306} (\bibinfo {year} {2019})}\BibitemShut {NoStop}%
\bibitem [{\citenamefont {Latt}\ \emph {et~al.}(2019)\citenamefont {Latt},
  \citenamefont {Coreixas}, \citenamefont {Beny},\ and\ \citenamefont
  {Parmigiani}}]{Latt2019}%
  \BibitemOpen
  \bibfield  {author} {\bibinfo {author} {\bibfnamefont {J.}~\bibnamefont
  {Latt}}, \bibinfo {author} {\bibfnamefont {C.}~\bibnamefont {Coreixas}},
  \bibinfo {author} {\bibfnamefont {J.}~\bibnamefont {Beny}}, \ and\ \bibinfo
  {author} {\bibfnamefont {A.}~\bibnamefont {Parmigiani}},\ }\href@noop {}
  {\enquote {\bibinfo {title} {Efficient supersonic flows through high-order
  guided equilibrium with lattice boltzmann},}\ } (\bibinfo {year} {2019}),\
  \Eprint {http://arxiv.org/abs/1910.13515} {arXiv:1910.13515
  [physics.comp-ph]} \BibitemShut {NoStop}%
\bibitem [{\citenamefont {Chen}(1998)}]{Chen1998}%
  \BibitemOpen
  \bibfield  {author} {\bibinfo {author} {\bibfnamefont {H.}~\bibnamefont
  {Chen}},\ }\href {\doibase 10.1103/PhysRevE.58.3955} {\bibfield  {journal}
  {\bibinfo  {journal} {Phys. Rev. E - Stat. Physics, Plasmas, Fluids, Relat.
  Interdiscip. Top.}\ }\textbf {\bibinfo {volume} {58}},\ \bibinfo {pages}
  {3955} (\bibinfo {year} {1998})}\BibitemShut {NoStop}%
\bibitem [{\citenamefont {Zhang}\ \emph {et~al.}(2001)\citenamefont {Zhang},
  \citenamefont {Chen}, \citenamefont {Qian},\ and\ \citenamefont
  {Chen}}]{Zhang2001}%
  \BibitemOpen
  \bibfield  {author} {\bibinfo {author} {\bibfnamefont {R.}~\bibnamefont
  {Zhang}}, \bibinfo {author} {\bibfnamefont {H.}~\bibnamefont {Chen}},
  \bibinfo {author} {\bibfnamefont {Y.}~\bibnamefont {Qian}}, \ and\ \bibinfo
  {author} {\bibfnamefont {S.}~\bibnamefont {Chen}},\ }\href {\doibase
  10.1103/PhysRevE.63.056705 M4 - Citavi} {\bibfield  {journal} {\bibinfo
  {journal} {Phys. Rev. E}\ }\textbf {\bibinfo {volume} {63}} (\bibinfo {year}
  {2001}),\ 10.1103/PhysRevE.63.056705 M4 - Citavi}\BibitemShut {NoStop}%
\bibitem [{\citenamefont {Guo}\ \emph {et~al.}(2015)\citenamefont {Guo},
  \citenamefont {Wang},\ and\ \citenamefont {Xu}}]{Guo2015}%
  \BibitemOpen
  \bibfield  {author} {\bibinfo {author} {\bibfnamefont {Z.}~\bibnamefont
  {Guo}}, \bibinfo {author} {\bibfnamefont {R.}~\bibnamefont {Wang}}, \ and\
  \bibinfo {author} {\bibfnamefont {K.}~\bibnamefont {Xu}},\ }\href {\doibase
  10.1103/PhysRevE.91.033313} {\bibfield  {journal} {\bibinfo  {journal} {Phys.
  Rev. E}\ }\textbf {\bibinfo {volume} {91}},\ \bibinfo {pages} {033313}
  (\bibinfo {year} {2015})}\BibitemShut {NoStop}%
\bibitem [{\citenamefont {Feng}\ \emph {et~al.}(2016)\citenamefont {Feng},
  \citenamefont {Sagaut},\ and\ \citenamefont {Tao}}]{Feng2016}%
  \BibitemOpen
  \bibfield  {author} {\bibinfo {author} {\bibfnamefont {Y.}~\bibnamefont
  {Feng}}, \bibinfo {author} {\bibfnamefont {P.}~\bibnamefont {Sagaut}}, \ and\
  \bibinfo {author} {\bibfnamefont {W.~Q.}\ \bibnamefont {Tao}},\ }\href
  {\doibase 10.1016/j.compfluid.2016.03.009} {\bibfield  {journal} {\bibinfo
  {journal} {Computers and Fluids}\ }\textbf {\bibinfo {volume} {131}},\
  \bibinfo {pages} {45} (\bibinfo {year} {2016})}\BibitemShut {NoStop}%
\bibitem [{\citenamefont {Shi}\ \emph {et~al.}(2001)\citenamefont {Shi},
  \citenamefont {Shyy},\ and\ \citenamefont {Mei}}]{Shi2001}%
  \BibitemOpen
  \bibfield  {author} {\bibinfo {author} {\bibfnamefont {W.}~\bibnamefont
  {Shi}}, \bibinfo {author} {\bibfnamefont {W.}~\bibnamefont {Shyy}}, \ and\
  \bibinfo {author} {\bibfnamefont {R.}~\bibnamefont {Mei}},\ }\href {\doibase
  10.1080/104077901300233578} {\bibfield  {journal} {\bibinfo  {journal}
  {Numerical Heat Transfer, Part B: Fundamentals}\ }\textbf {\bibinfo {volume}
  {40}},\ \bibinfo {pages} {1} (\bibinfo {year} {2001})}\BibitemShut {NoStop}%
\bibitem [{\citenamefont {Wang}\ \emph {et~al.}(2007)\citenamefont {Wang},
  \citenamefont {He}, \citenamefont {Zhao}, \citenamefont {Tang},\ and\
  \citenamefont {Tao}}]{WANG2007}%
  \BibitemOpen
  \bibfield  {author} {\bibinfo {author} {\bibfnamefont {Y.}~\bibnamefont
  {Wang}}, \bibinfo {author} {\bibfnamefont {Y.~L.}\ \bibnamefont {He}},
  \bibinfo {author} {\bibfnamefont {T.~S.}\ \bibnamefont {Zhao}}, \bibinfo
  {author} {\bibfnamefont {G.~H.}\ \bibnamefont {Tang}}, \ and\ \bibinfo
  {author} {\bibfnamefont {W.~Q.}\ \bibnamefont {Tao}},\ }\href {\doibase
  10.1142/S0129183107011868} {\bibfield  {journal} {\bibinfo  {journal}
  {International Journal of Modern Physics C}\ }\textbf {\bibinfo {volume}
  {18}},\ \bibinfo {pages} {1961} (\bibinfo {year} {2007})}\BibitemShut
  {NoStop}%
\bibitem [{\citenamefont {Nie}\ \emph {et~al.}(2009)\citenamefont {Nie},
  \citenamefont {Shan},\ and\ \citenamefont {Chen}}]{Nie2009}%
  \BibitemOpen
  \bibfield  {author} {\bibinfo {author} {\bibfnamefont {X.}~\bibnamefont
  {Nie}}, \bibinfo {author} {\bibfnamefont {X.}~\bibnamefont {Shan}}, \ and\
  \bibinfo {author} {\bibfnamefont {H.}~\bibnamefont {Chen}},\ }\href {\doibase
  10.2514/6.2009-139} {\bibfield  {journal} {\bibinfo  {journal} {47th AIAA
  Aerospace Sciences Meeting including The New Horizons Forum and Aerospace
  Exposition}\ } (\bibinfo {year} {2009}),\ 10.2514/6.2009-139}\BibitemShut
  {NoStop}%
\bibitem [{\citenamefont {Frapolli}(2017)}]{Frapolli2017}%
  \BibitemOpen
  \bibfield  {author} {\bibinfo {author} {\bibfnamefont {N.}~\bibnamefont
  {Frapolli}},\ }\emph {\bibinfo {title} {{Entropic lattice Boltzmann models
  for thermal and compressible flows}}},\ \href {\doibase
  10.3929/ETHZ-A-010890892} {Ph.D. thesis} (\bibinfo {year} {2017})\BibitemShut
  {NoStop}%
\bibitem [{\citenamefont {Kataoka}\ and\ \citenamefont
  {Tsutahara}(2004)}]{Kataoka2004}%
  \BibitemOpen
  \bibfield  {author} {\bibinfo {author} {\bibfnamefont {T.}~\bibnamefont
  {Kataoka}}\ and\ \bibinfo {author} {\bibfnamefont {M.}~\bibnamefont
  {Tsutahara}},\ }\href {\doibase 10.1103/PhysRevE.69.056702} {\bibfield
  {journal} {\bibinfo  {journal} {Phys. Rev. E}\ } (\bibinfo {year} {2004}),\
  10.1103/PhysRevE.69.056702}\BibitemShut {NoStop}%
\bibitem [{\citenamefont {Li}\ \emph {et~al.}(2007)\citenamefont {Li},
  \citenamefont {He}, \citenamefont {Wang},\ and\ \citenamefont
  {Tao}}]{Li2007}%
  \BibitemOpen
  \bibfield  {author} {\bibinfo {author} {\bibfnamefont {Q.}~\bibnamefont
  {Li}}, \bibinfo {author} {\bibfnamefont {Y.~L.}\ \bibnamefont {He}}, \bibinfo
  {author} {\bibfnamefont {Y.}~\bibnamefont {Wang}}, \ and\ \bibinfo {author}
  {\bibfnamefont {W.~Q.}\ \bibnamefont {Tao}},\ }\href {\doibase
  10.1103/PhysRevE.76.056705} {\bibfield  {journal} {\bibinfo  {journal} {Phys.
  Rev. E}\ } (\bibinfo {year} {2007}),\ 10.1103/PhysRevE.76.056705}\BibitemShut
  {NoStop}%
\bibitem [{\citenamefont {Coreixas}(2018)}]{Coreixas2018}%
  \BibitemOpen
  \bibfield  {author} {\bibinfo {author} {\bibfnamefont {C.}~\bibnamefont
  {Coreixas}},\ }\emph {\bibinfo {title} {{High-order extension of the
  recursive regularized lattice Boltzmann method}}},\ \href@noop {} {Ph.D.
  thesis} (\bibinfo {year} {2018})\BibitemShut {NoStop}%
\bibitem [{\citenamefont {Chikatamarla}\ and\ \citenamefont
  {Karlin}(2009)}]{Chikatamarla2009}%
  \BibitemOpen
  \bibfield  {author} {\bibinfo {author} {\bibfnamefont {S.~S.}\ \bibnamefont
  {Chikatamarla}}\ and\ \bibinfo {author} {\bibfnamefont {I.~V.}\ \bibnamefont
  {Karlin}},\ }\href {\doibase 10.1103/PhysRevE.79.046701} {\bibfield
  {journal} {\bibinfo  {journal} {Phys. Rev. E}\ }\textbf {\bibinfo {volume}
  {79}} (\bibinfo {year} {2009}),\ 10.1103/PhysRevE.79.046701}\BibitemShut
  {NoStop}%
\bibitem [{\citenamefont {Frapolli}\ \emph {et~al.}(2016)\citenamefont
  {Frapolli}, \citenamefont {Chikatamarla},\ and\ \citenamefont
  {Karlin}}]{Frapolli2016}%
  \BibitemOpen
  \bibfield  {author} {\bibinfo {author} {\bibfnamefont {N.}~\bibnamefont
  {Frapolli}}, \bibinfo {author} {\bibfnamefont {S.~S.}\ \bibnamefont
  {Chikatamarla}}, \ and\ \bibinfo {author} {\bibfnamefont {I.~V.}\
  \bibnamefont {Karlin}},\ }\href {\doibase 10.1103/PhysRevE.93.063302}
  {\bibfield  {journal} {\bibinfo  {journal} {Phys. Rev. E}\ }\textbf {\bibinfo
  {volume} {93}},\ \bibinfo {pages} {063302} (\bibinfo {year}
  {2016})}\BibitemShut {NoStop}%
\bibitem [{\citenamefont {Coreixas}\ \emph {et~al.}(2017)\citenamefont
  {Coreixas}, \citenamefont {Wissocq}, \citenamefont {Puigt}, \citenamefont
  {Boussuge},\ and\ \citenamefont {Sagaut}}]{Coreixas2017}%
  \BibitemOpen
  \bibfield  {author} {\bibinfo {author} {\bibfnamefont {C.}~\bibnamefont
  {Coreixas}}, \bibinfo {author} {\bibfnamefont {G.}~\bibnamefont {Wissocq}},
  \bibinfo {author} {\bibfnamefont {G.}~\bibnamefont {Puigt}}, \bibinfo
  {author} {\bibfnamefont {J.~F.}\ \bibnamefont {Boussuge}}, \ and\ \bibinfo
  {author} {\bibfnamefont {P.}~\bibnamefont {Sagaut}},\ }\href {\doibase
  10.1103/PhysRevE.96.033306} {\bibfield  {journal} {\bibinfo  {journal} {Phys.
  Rev. E}\ } (\bibinfo {year} {2017}),\ 10.1103/PhysRevE.96.033306},\ \Eprint
  {http://arxiv.org/abs/1704.04413} {arXiv:1704.04413} \BibitemShut {NoStop}%
\bibitem [{\citenamefont {Kr{\"{a}}mer}\ \emph {et~al.}(2017)\citenamefont
  {Kr{\"{a}}mer}, \citenamefont {K{\"{u}}llmer}, \citenamefont {Reith},
  \citenamefont {Joppich},\ and\ \citenamefont {Foysi}}]{Kramer2017a}%
  \BibitemOpen
  \bibfield  {author} {\bibinfo {author} {\bibfnamefont {A.}~\bibnamefont
  {Kr{\"{a}}mer}}, \bibinfo {author} {\bibfnamefont {K.}~\bibnamefont
  {K{\"{u}}llmer}}, \bibinfo {author} {\bibfnamefont {D.}~\bibnamefont
  {Reith}}, \bibinfo {author} {\bibfnamefont {W.}~\bibnamefont {Joppich}}, \
  and\ \bibinfo {author} {\bibfnamefont {H.}~\bibnamefont {Foysi}},\ }\href
  {\doibase 10.1103/PhysRevE.95.023305} {\bibfield  {journal} {\bibinfo
  {journal} {Phys. Rev. E}\ }\textbf {\bibinfo {volume} {95}} (\bibinfo {year}
  {2017}),\ 10.1103/PhysRevE.95.023305}\BibitemShut {NoStop}%
\bibitem [{\citenamefont {Kr{\"{a}}mer}(2017)}]{Kramer}%
  \BibitemOpen
  \bibfield  {author} {\bibinfo {author} {\bibfnamefont {A.}~\bibnamefont
  {Kr{\"{a}}mer}},\ }\emph {\bibinfo {title} {{Lattice-Boltzmann-Methoden zur
  Simulation inkompressibler Wirbelstr{\"{o}}mungen}}},\ \href
  {http://dokumentix.ub.uni-siegen.de/opus/volltexte/2017/1237/} {Ph.D. thesis}
  (\bibinfo {year} {2017})\BibitemShut {NoStop}%
\bibitem [{\citenamefont {{Di Ilio}}\ \emph {et~al.}(2018)\citenamefont {{Di
  Ilio}}, \citenamefont {Dorschner}, \citenamefont {Bella}, \citenamefont
  {Succi},\ and\ \citenamefont {Karlin}}]{DiIlio2018}%
  \BibitemOpen
  \bibfield  {author} {\bibinfo {author} {\bibfnamefont {G.}~\bibnamefont {{Di
  Ilio}}}, \bibinfo {author} {\bibfnamefont {B.}~\bibnamefont {Dorschner}},
  \bibinfo {author} {\bibfnamefont {G.}~\bibnamefont {Bella}}, \bibinfo
  {author} {\bibfnamefont {S.}~\bibnamefont {Succi}}, \ and\ \bibinfo {author}
  {\bibfnamefont {I.~V.}\ \bibnamefont {Karlin}},\ }\href {\doibase
  10.1017/jfm.2018.413} {\bibfield  {journal} {\bibinfo  {journal} {J. Fluid
  Mech.}\ }\textbf {\bibinfo {volume} {849}},\ \bibinfo {pages} {35} (\bibinfo
  {year} {2018})}\BibitemShut {NoStop}%
\bibitem [{\citenamefont {Kr{\"{a}}mer}\ \emph {et~al.}(2020)\citenamefont
  {Kr{\"{a}}mer}, \citenamefont {Wilde}, \citenamefont {K{\"{u}}llmer},
  \citenamefont {Reith}, \citenamefont {Foysi},\ and\ \citenamefont
  {Joppich}}]{Kramer2018}%
  \BibitemOpen
  \bibfield  {author} {\bibinfo {author} {\bibfnamefont {A.}~\bibnamefont
  {Kr{\"{a}}mer}}, \bibinfo {author} {\bibfnamefont {D.}~\bibnamefont {Wilde}},
  \bibinfo {author} {\bibfnamefont {K.}~\bibnamefont {K{\"{u}}llmer}}, \bibinfo
  {author} {\bibfnamefont {D.}~\bibnamefont {Reith}}, \bibinfo {author}
  {\bibfnamefont {H.}~\bibnamefont {Foysi}}, \ and\ \bibinfo {author}
  {\bibfnamefont {W.}~\bibnamefont {Joppich}},\ }\href {\doibase
  10.1016/j.camwa.2018.10.041} {\bibfield  {journal} {\bibinfo  {journal}
  {Computers and Mathematics with Applications}\ }\textbf {\bibinfo {volume}
  {79}},\ \bibinfo {pages} {34} (\bibinfo {year} {2020})}\BibitemShut {NoStop}%
\bibitem [{\citenamefont {Zakirov}\ \emph {et~al.}(2019)\citenamefont
  {Zakirov}, \citenamefont {Korneev}, \citenamefont {Levchenko},\ and\
  \citenamefont {Perepelkina}}]{Zarikov2019}%
  \BibitemOpen
  \bibfield  {author} {\bibinfo {author} {\bibfnamefont {A.~V.}\ \bibnamefont
  {Zakirov}}, \bibinfo {author} {\bibfnamefont {B.~A.}\ \bibnamefont
  {Korneev}}, \bibinfo {author} {\bibfnamefont {V.~D.}\ \bibnamefont
  {Levchenko}}, \ and\ \bibinfo {author} {\bibfnamefont {A.~Y.}\ \bibnamefont
  {Perepelkina}},\ }\href {\doibase 10.20948/prepr-2019-35-e} {\bibfield
  {journal} {\bibinfo  {journal} {Keldysh Inst. Prepr.}\ ,\ \bibinfo {pages}
  {1}} (\bibinfo {year} {2019})}\BibitemShut {NoStop}%
\bibitem [{\citenamefont {Sun}(1998)}]{Sun1998}%
  \BibitemOpen
  \bibfield  {author} {\bibinfo {author} {\bibfnamefont {C.}~\bibnamefont
  {Sun}},\ }\href {\doibase 10.1103/PhysRevE.58.7283} {\bibfield  {journal}
  {\bibinfo  {journal} {Phys. Rev. E - Stat. Physics, Plasmas, Fluids, Relat.
  Interdiscip. Top.}\ }\textbf {\bibinfo {volume} {58}},\ \bibinfo {pages}
  {7283} (\bibinfo {year} {1998})}\BibitemShut {NoStop}%
\bibitem [{\citenamefont {Sun}(2000)}]{Sun2000}%
  \BibitemOpen
  \bibfield  {author} {\bibinfo {author} {\bibfnamefont {C.}~\bibnamefont
  {Sun}},\ }\href {\doibase 10.1006/jcph.2000.6487} {\bibfield  {journal}
  {\bibinfo  {journal} {J. Comput. Phys.}\ }\textbf {\bibinfo {volume} {161}},\
  \bibinfo {pages} {70} (\bibinfo {year} {2000})}\BibitemShut {NoStop}%
\bibitem [{\citenamefont {Sun}\ and\ \citenamefont {Hsu}(2003)}]{Sun2003}%
  \BibitemOpen
  \bibfield  {author} {\bibinfo {author} {\bibfnamefont {C.}~\bibnamefont
  {Sun}}\ and\ \bibinfo {author} {\bibfnamefont {A.~T.}\ \bibnamefont {Hsu}},\
  }\href {\doibase 10.1103/PhysRevE.68.016303} {\bibfield  {journal} {\bibinfo
  {journal} {Phys. Rev. E}\ }\textbf {\bibinfo {volume} {68}},\ \bibinfo
  {pages} {14} (\bibinfo {year} {2003})}\BibitemShut {NoStop}%
\bibitem [{\citenamefont {Shan}\ and\ \citenamefont {He}(1998)}]{Shan1998}%
  \BibitemOpen
  \bibfield  {author} {\bibinfo {author} {\bibfnamefont {X.}~\bibnamefont
  {Shan}}\ and\ \bibinfo {author} {\bibfnamefont {X.}~\bibnamefont {He}},\
  }\href {\doibase 10.1103/PhysRevLett.80.65} {\bibfield  {journal} {\bibinfo
  {journal} {Phys. Rev. Lett.}\ }\textbf {\bibinfo {volume} {80}},\ \bibinfo
  {pages} {65} (\bibinfo {year} {1998})},\ \Eprint
  {http://arxiv.org/abs/9712001} {9712001 [comp-gas]} \BibitemShut {NoStop}%
\bibitem [{\citenamefont {Shan}\ \emph {et~al.}(2006)\citenamefont {Shan},
  \citenamefont {Yuan},\ and\ \citenamefont {Chen}}]{Shan2006}%
  \BibitemOpen
  \bibfield  {author} {\bibinfo {author} {\bibfnamefont {X.}~\bibnamefont
  {Shan}}, \bibinfo {author} {\bibfnamefont {X.~F.}\ \bibnamefont {Yuan}}, \
  and\ \bibinfo {author} {\bibfnamefont {H.}~\bibnamefont {Chen}},\ }\href
  {\doibase 10.1017/S0022112005008153} {\bibfield  {journal} {\bibinfo
  {journal} {Journal of Fluid Mechanics}\ } (\bibinfo {year} {2006}),\
  10.1017/S0022112005008153}\BibitemShut {NoStop}%
\bibitem [{\citenamefont {Bardow}\ \emph {et~al.}(2008)\citenamefont {Bardow},
  \citenamefont {Karlin},\ and\ \citenamefont {Gusev}}]{Bardow2008}%
  \BibitemOpen
  \bibfield  {author} {\bibinfo {author} {\bibfnamefont {A.}~\bibnamefont
  {Bardow}}, \bibinfo {author} {\bibfnamefont {I.~V.}\ \bibnamefont {Karlin}},
  \ and\ \bibinfo {author} {\bibfnamefont {A.~A.}\ \bibnamefont {Gusev}},\
  }\href {\doibase 10.1103/PhysRevE.77.025701} {\bibfield  {journal} {\bibinfo
  {journal} {Phys. Rev. E}\ } (\bibinfo {year} {2008}),\
  10.1103/PhysRevE.77.025701}\BibitemShut {NoStop}%
\bibitem [{\citenamefont {Shan}(2016)}]{Shan2016}%
  \BibitemOpen
  \bibfield  {author} {\bibinfo {author} {\bibfnamefont {X.}~\bibnamefont
  {Shan}},\ }\href {\doibase 10.1016/j.jocs.2016.03.002} {\bibfield  {journal}
  {\bibinfo  {journal} {Journal of Computational Science}\ }\textbf {\bibinfo
  {volume} {17}},\ \bibinfo {pages} {475} (\bibinfo {year} {2016})}\BibitemShut
  {NoStop}%
\bibitem [{\citenamefont {Grad}(1949)}]{Grad1949}%
  \BibitemOpen
  \bibfield  {author} {\bibinfo {author} {\bibfnamefont {H.}~\bibnamefont
  {Grad}},\ }\href {\doibase 10.1002/cpa.3160020403} {\bibfield  {journal}
  {\bibinfo  {journal} {Communications on Pure and Applied Mathematics}\
  }\textbf {\bibinfo {volume} {2}},\ \bibinfo {pages} {331} (\bibinfo {year}
  {1949})}\BibitemShut {NoStop}%
\bibitem [{\citenamefont {Malaspinas}(2009)}]{Malaspinas2009}%
  \BibitemOpen
  \bibfield  {author} {\bibinfo {author} {\bibfnamefont {O.}~\bibnamefont
  {Malaspinas}},\ }\emph {\bibinfo {title} {{Lattice Boltzmann Method for the
  Simulation of Viscoelastic Fluid Flows}}},\ \href@noop {} {Ph.D. thesis}
  (\bibinfo {year} {2009})\BibitemShut {NoStop}%
\bibitem [{\citenamefont {Bhatnagar}\ \emph {et~al.}(1954)\citenamefont
  {Bhatnagar}, \citenamefont {Gross},\ and\ \citenamefont
  {Krook}}]{Bhatnagar1954}%
  \BibitemOpen
  \bibfield  {author} {\bibinfo {author} {\bibfnamefont {P.~L.}\ \bibnamefont
  {Bhatnagar}}, \bibinfo {author} {\bibfnamefont {E.~P.}\ \bibnamefont
  {Gross}}, \ and\ \bibinfo {author} {\bibfnamefont {M.}~\bibnamefont
  {Krook}},\ }\href {\doibase 10.1103/PhysRev.94.511} {\bibfield  {journal}
  {\bibinfo  {journal} {Physical Review}\ }\textbf {\bibinfo {volume} {94}},\
  \bibinfo {pages} {511} (\bibinfo {year} {1954})}\BibitemShut {NoStop}%
\bibitem [{\citenamefont {Suchy}(1996)}]{Suchy1996}%
  \BibitemOpen
  \bibfield  {author} {\bibinfo {author} {\bibfnamefont {K.}~\bibnamefont
  {Suchy}},\ }\href@noop {} {\bibfield  {journal} {\bibinfo  {journal}
  {Australian J. of Physics}\ }\textbf {\bibinfo {volume} {49(6)}} (\bibinfo
  {year} {1996})}\BibitemShut {NoStop}%
\bibitem [{\citenamefont {Nie}\ \emph {et~al.}(2008{\natexlab{a}})\citenamefont
  {Nie}, \citenamefont {Shan},\ and\ \citenamefont {Chen}}]{Nie2008a}%
  \BibitemOpen
  \bibfield  {author} {\bibinfo {author} {\bibfnamefont {X.~B.}\ \bibnamefont
  {Nie}}, \bibinfo {author} {\bibfnamefont {X.}~\bibnamefont {Shan}}, \ and\
  \bibinfo {author} {\bibfnamefont {H.}~\bibnamefont {Chen}},\ }\href {\doibase
  10.1209/0295-5075/81/34005} {\bibfield  {journal} {\bibinfo  {journal}
  {Europhys. Lett.}\ }\textbf {\bibinfo {volume} {81}},\ \bibinfo {pages}
  {34005} (\bibinfo {year} {2008}{\natexlab{a}})}\BibitemShut {NoStop}%
\bibitem [{\citenamefont {Shan}\ and\ \citenamefont {Chen}(2007)}]{Shan2007}%
  \BibitemOpen
  \bibfield  {author} {\bibinfo {author} {\bibfnamefont {X.}~\bibnamefont
  {Shan}}\ and\ \bibinfo {author} {\bibfnamefont {H.}~\bibnamefont {Chen}},\
  }\href {\doibase 10.1142/S0129183107010887} {\bibfield  {journal} {\bibinfo
  {journal} {International Journal of Modern Physics C}\ }\textbf {\bibinfo
  {volume} {18}},\ \bibinfo {pages} {635} (\bibinfo {year} {2007})}\BibitemShut
  {NoStop}%
\bibitem [{\citenamefont {Ansumali}\ \emph {et~al.}(2007)\citenamefont
  {Ansumali}, \citenamefont {Arcidiacono}, \citenamefont {Chikatamarla},
  \citenamefont {Prasianakis}, \citenamefont {Gorban},\ and\ \citenamefont
  {Karlin}}]{Ansumali2007}%
  \BibitemOpen
  \bibfield  {author} {\bibinfo {author} {\bibfnamefont {S.}~\bibnamefont
  {Ansumali}}, \bibinfo {author} {\bibfnamefont {S.}~\bibnamefont
  {Arcidiacono}}, \bibinfo {author} {\bibfnamefont {S.~S.}\ \bibnamefont
  {Chikatamarla}}, \bibinfo {author} {\bibfnamefont {N.~I.}\ \bibnamefont
  {Prasianakis}}, \bibinfo {author} {\bibfnamefont {A.~N.}\ \bibnamefont
  {Gorban}}, \ and\ \bibinfo {author} {\bibfnamefont {I.~V.}\ \bibnamefont
  {Karlin}},\ }\href {\doibase 10.1140/epjb/e2007-00100-1} {\bibfield
  {journal} {\bibinfo  {journal} {The European Physical Journal B}\ }\textbf
  {\bibinfo {volume} {56}},\ \bibinfo {pages} {135} (\bibinfo {year}
  {2007})}\BibitemShut {NoStop}%
\bibitem [{\citenamefont {Shakhov}(1972)}]{Shakhov1972}%
  \BibitemOpen
  \bibfield  {author} {\bibinfo {author} {\bibfnamefont {E.~M.}\ \bibnamefont
  {Shakhov}},\ }\href {\doibase 10.1007/BF01029546} {\bibfield  {journal}
  {\bibinfo  {journal} {Fluid Dynamics}\ }\textbf {\bibinfo {volume} {3}},\
  \bibinfo {pages} {95} (\bibinfo {year} {1972})}\BibitemShut {NoStop}%
\bibitem [{\citenamefont {Nie}\ \emph {et~al.}(2008{\natexlab{b}})\citenamefont
  {Nie}, \citenamefont {Shan},\ and\ \citenamefont {Chen}}]{Nie2008}%
  \BibitemOpen
  \bibfield  {author} {\bibinfo {author} {\bibfnamefont {X.}~\bibnamefont
  {Nie}}, \bibinfo {author} {\bibfnamefont {X.}~\bibnamefont {Shan}}, \ and\
  \bibinfo {author} {\bibfnamefont {H.}~\bibnamefont {Chen}},\ }\href {\doibase
  10.1103/PhysRevE.77.035701} {\bibfield  {journal} {\bibinfo  {journal} {Phys.
  Rev. E}\ }\textbf {\bibinfo {volume} {77}},\ \bibinfo {pages} {1} (\bibinfo
  {year} {2008}{\natexlab{b}})}\BibitemShut {NoStop}%
\bibitem [{\citenamefont {Rykov}(1976)}]{Rykov1976}%
  \BibitemOpen
  \bibfield  {author} {\bibinfo {author} {\bibfnamefont {V.~A.}\ \bibnamefont
  {Rykov}},\ }\href {\doibase 10.1007/BF01023275} {\bibfield  {journal}
  {\bibinfo  {journal} {Fluid Dynamics}\ }\textbf {\bibinfo {volume} {10}},\
  \bibinfo {pages} {959} (\bibinfo {year} {1976})}\BibitemShut {NoStop}%
\bibitem [{\citenamefont {Dellar}(2008)}]{Dellar2008}%
  \BibitemOpen
  \bibfield  {author} {\bibinfo {author} {\bibfnamefont {P.~J.}\ \bibnamefont
  {Dellar}},\ }\href {\doibase 10.1504/PCFD.2008.018081} {\bibfield  {journal}
  {\bibinfo  {journal} {Progress in Computational Fluid Dynamics, An
  International Journal}\ } (\bibinfo {year} {2008}),\
  10.1504/PCFD.2008.018081}\BibitemShut {NoStop}%
\bibitem [{\citenamefont {He}\ \emph {et~al.}(1998)\citenamefont {He},
  \citenamefont {Shan},\ and\ \citenamefont {Doolen}}]{He1998a}%
  \BibitemOpen
  \bibfield  {author} {\bibinfo {author} {\bibfnamefont {X.}~\bibnamefont
  {He}}, \bibinfo {author} {\bibfnamefont {X.}~\bibnamefont {Shan}}, \ and\
  \bibinfo {author} {\bibfnamefont {G.~D.}\ \bibnamefont {Doolen}},\ }\href
  {\doibase 10.1103/PhysRevE.57.R13} {\bibfield  {journal} {\bibinfo  {journal}
  {Phys. Rev. E - Stat. Physics, Plasmas, Fluids, Relat. Interdiscip. Top.}\
  }\textbf {\bibinfo {volume} {57}},\ \bibinfo {pages} {R13} (\bibinfo {year}
  {1998})}\BibitemShut {NoStop}%
\bibitem [{\citenamefont {Krause}(2010)}]{Krause2010}%
  \BibitemOpen
  \bibfield  {author} {\bibinfo {author} {\bibfnamefont {M.}~\bibnamefont
  {Krause}},\ }\emph {\bibinfo {title} {Fluid Flow Simulation and Optimisation
  with Lattice Boltzmann Methods on High Performance Computers - Application to
  the Human Respiratory System}},\ \href@noop {} {Ph.D. thesis} (\bibinfo
  {year} {2010})\BibitemShut {NoStop}%
\bibitem [{\citenamefont {Wilde}\ \emph {et~al.}(2019)\citenamefont {Wilde},
  \citenamefont {Kr{\"{a}}mer}, \citenamefont {K{\"{u}}llmer}, \citenamefont
  {Foysi},\ and\ \citenamefont {Reith}}]{Wilde2019}%
  \BibitemOpen
  \bibfield  {author} {\bibinfo {author} {\bibfnamefont {D.}~\bibnamefont
  {Wilde}}, \bibinfo {author} {\bibfnamefont {A.}~\bibnamefont {Kr{\"{a}}mer}},
  \bibinfo {author} {\bibfnamefont {K.}~\bibnamefont {K{\"{u}}llmer}}, \bibinfo
  {author} {\bibfnamefont {H.}~\bibnamefont {Foysi}}, \ and\ \bibinfo {author}
  {\bibfnamefont {D.}~\bibnamefont {Reith}},\ }\href {\doibase
  10.1002/fld.4716} {\bibfield  {journal} {\bibinfo  {journal} {Int. J. Numer.
  Methods Fluids}\ }\textbf {\bibinfo {volume} {90}},\ \bibinfo {pages} {156}
  (\bibinfo {year} {2019})}\BibitemShut {NoStop}%
\bibitem [{\citenamefont {Hesthaven}\ and\ \citenamefont
  {Warburton}(2002)}]{Hesthaven2002}%
  \BibitemOpen
  \bibfield  {author} {\bibinfo {author} {\bibfnamefont {J.}~\bibnamefont
  {Hesthaven}}\ and\ \bibinfo {author} {\bibfnamefont {T.}~\bibnamefont
  {Warburton}},\ }\href {\doibase 10.1006/jcph.2002.7118} {\bibfield  {journal}
  {\bibinfo  {journal} {J. Comput. Phys.}\ } (\bibinfo {year} {2002}),\
  10.1006/jcph.2002.7118}\BibitemShut {NoStop}%
\bibitem [{\citenamefont {Einkemmer}\ and\ \citenamefont
  {Ostermann}(2015)}]{Einkemmer2015}%
  \BibitemOpen
  \bibfield  {author} {\bibinfo {author} {\bibfnamefont {L.}~\bibnamefont
  {Einkemmer}}\ and\ \bibinfo {author} {\bibfnamefont {A.}~\bibnamefont
  {Ostermann}},\ }\href {\doibase 10.1016/j.camwa.2014.12.004 PM - 25844018 M4
  - Citavi} {\bibfield  {journal} {\bibinfo  {journal} {Computers {\&}
  mathematics with applications (Oxford, England : 1987)}\ }\textbf {\bibinfo
  {volume} {69}},\ \bibinfo {pages} {170} (\bibinfo {year} {2015})}\BibitemShut
  {NoStop}%
\bibitem [{\citenamefont {Falcone}\ and\ \citenamefont
  {Ferretti}(2013)}]{Falcone2013}%
  \BibitemOpen
  \bibfield  {author} {\bibinfo {author} {\bibfnamefont {M.}~\bibnamefont
  {Falcone}}\ and\ \bibinfo {author} {\bibfnamefont {R.}~\bibnamefont
  {Ferretti}},\ }\href {\doibase 10.1137/1.9781611973051} {\emph {\bibinfo
  {title} {Semi-Lagrangian Approximation Schemes for Linear and
  Hamilton—Jacobi Equations}}}\ (\bibinfo  {publisher} {Society for
  Industrial and Applied Mathematics},\ \bibinfo {address} {Philadelphia, PA},\
  \bibinfo {year} {2013})\ \Eprint
  {http://arxiv.org/abs/https://epubs.siam.org/doi/pdf/10.1137/1.9781611973051}
  {https://epubs.siam.org/doi/pdf/10.1137/1.9781611973051} \BibitemShut
  {NoStop}%
\bibitem [{\citenamefont {Bangerth}\ \emph {et~al.}(2007)\citenamefont
  {Bangerth}, \citenamefont {Hartmann},\ and\ \citenamefont
  {Kanschat}}]{Bangerth2007}%
  \BibitemOpen
  \bibfield  {author} {\bibinfo {author} {\bibfnamefont {W.}~\bibnamefont
  {Bangerth}}, \bibinfo {author} {\bibfnamefont {R.}~\bibnamefont {Hartmann}},
  \ and\ \bibinfo {author} {\bibfnamefont {G.}~\bibnamefont {Kanschat}},\
  }\href {\doibase 10.1145/1268776.1268779} {\bibfield  {journal} {\bibinfo
  {journal} {ACM Transactions on Mathematical Software}\ } (\bibinfo {year}
  {2007}),\ 10.1145/1268776.1268779}\BibitemShut {NoStop}%
\bibitem [{\citenamefont {Philippi}\ \emph {et~al.}(2006)\citenamefont
  {Philippi}, \citenamefont {Hegele}, \citenamefont {{Dos Santos}},\ and\
  \citenamefont {Surmas}}]{Philippi2006}%
  \BibitemOpen
  \bibfield  {author} {\bibinfo {author} {\bibfnamefont {P.~C.}\ \bibnamefont
  {Philippi}}, \bibinfo {author} {\bibfnamefont {L.~A.}\ \bibnamefont
  {Hegele}}, \bibinfo {author} {\bibfnamefont {L.~O.~E.}\ \bibnamefont {{Dos
  Santos}}}, \ and\ \bibinfo {author} {\bibfnamefont {R.}~\bibnamefont
  {Surmas}},\ }\href {\doibase 10.1103/PhysRevE.73.056702} {\bibfield
  {journal} {\bibinfo  {journal} {Phys. Rev. E - Stat. Nonlinear, Soft Matter
  Phys.}\ }\textbf {\bibinfo {volume} {73}} (\bibinfo {year} {2006}),\
  10.1103/PhysRevE.73.056702}\BibitemShut {NoStop}%
\bibitem [{\citenamefont {Kr{\"{a}}mer}(2019)}]{lettuce}%
  \BibitemOpen
  \bibfield  {author} {\bibinfo {author} {\bibfnamefont {A.}~\bibnamefont
  {Kr{\"{a}}mer}},\ }\href@noop {} {} (\bibinfo {year} {2019}),\ \bibinfo
  {note} {\url{https://github.com/Olllom/lettuce}}\BibitemShut {NoStop}%
\bibitem [{\citenamefont {Paszke}\ \emph {et~al.}(2019)\citenamefont {Paszke},
  \citenamefont {Gross}, \citenamefont {Massa}, \citenamefont {Lerer},
  \citenamefont {Bradbury}, \citenamefont {Chanan}, \citenamefont {Killeen},
  \citenamefont {Lin}, \citenamefont {Gimelshein}, \citenamefont {Antiga},
  \citenamefont {Desmaison}, \citenamefont {Kopf}, \citenamefont {Yang},
  \citenamefont {DeVito}, \citenamefont {Raison}, \citenamefont {Tejani},
  \citenamefont {Chilamkurthy}, \citenamefont {Steiner}, \citenamefont {Fang},
  \citenamefont {Bai},\ and\ \citenamefont {Chintala}}]{pytorch}%
  \BibitemOpen
  \bibfield  {author} {\bibinfo {author} {\bibfnamefont {A.}~\bibnamefont
  {Paszke}}, \bibinfo {author} {\bibfnamefont {S.}~\bibnamefont {Gross}},
  \bibinfo {author} {\bibfnamefont {F.}~\bibnamefont {Massa}}, \bibinfo
  {author} {\bibfnamefont {A.}~\bibnamefont {Lerer}}, \bibinfo {author}
  {\bibfnamefont {J.}~\bibnamefont {Bradbury}}, \bibinfo {author}
  {\bibfnamefont {G.}~\bibnamefont {Chanan}}, \bibinfo {author} {\bibfnamefont
  {T.}~\bibnamefont {Killeen}}, \bibinfo {author} {\bibfnamefont
  {Z.}~\bibnamefont {Lin}}, \bibinfo {author} {\bibfnamefont {N.}~\bibnamefont
  {Gimelshein}}, \bibinfo {author} {\bibfnamefont {L.}~\bibnamefont {Antiga}},
  \bibinfo {author} {\bibfnamefont {A.}~\bibnamefont {Desmaison}}, \bibinfo
  {author} {\bibfnamefont {A.}~\bibnamefont {Kopf}}, \bibinfo {author}
  {\bibfnamefont {E.}~\bibnamefont {Yang}}, \bibinfo {author} {\bibfnamefont
  {Z.}~\bibnamefont {DeVito}}, \bibinfo {author} {\bibfnamefont
  {M.}~\bibnamefont {Raison}}, \bibinfo {author} {\bibfnamefont
  {A.}~\bibnamefont {Tejani}}, \bibinfo {author} {\bibfnamefont
  {S.}~\bibnamefont {Chilamkurthy}}, \bibinfo {author} {\bibfnamefont
  {B.}~\bibnamefont {Steiner}}, \bibinfo {author} {\bibfnamefont
  {L.}~\bibnamefont {Fang}}, \bibinfo {author} {\bibfnamefont {J.}~\bibnamefont
  {Bai}}, \ and\ \bibinfo {author} {\bibfnamefont {S.}~\bibnamefont
  {Chintala}},\ }in\ \href
  {http://papers.neurips.cc/paper/9015-pytorch-an-imperative-style-high-performance-deep-learning-library.pdf}
  {\emph {\bibinfo {booktitle} {Advances in Neural Information Processing
  Systems 32}}},\ \bibinfo {editor} {edited by\ \bibinfo {editor}
  {\bibfnamefont {H.}~\bibnamefont {Wallach}}, \bibinfo {editor} {\bibfnamefont
  {H.}~\bibnamefont {Larochelle}}, \bibinfo {editor} {\bibfnamefont
  {A.}~\bibnamefont {Beygelzimer}}, \bibinfo {editor} {\bibfnamefont
  {F.}~\bibnamefont {d'Alch\'{e} Buc}}, \bibinfo {editor} {\bibfnamefont
  {E.}~\bibnamefont {Fox}}, \ and\ \bibinfo {editor} {\bibfnamefont
  {R.}~\bibnamefont {Garnett}}}\ (\bibinfo  {publisher} {Curran Associates,
  Inc.},\ \bibinfo {year} {2019})\ pp.\ \bibinfo {pages}
  {8024--8035}\BibitemShut {NoStop}%
\bibitem [{\citenamefont {{Sod}}(1978)}]{Sod1978}%
  \BibitemOpen
  \bibfield  {author} {\bibinfo {author} {\bibfnamefont {G.~A.}\ \bibnamefont
  {{Sod}}},\ }\href {\doibase 10.1016/0021-9991(78)90023-2} {\bibfield
  {journal} {\bibinfo  {journal} {Journal of Computational Physics}\ }\textbf
  {\bibinfo {volume} {27}},\ \bibinfo {pages} {1} (\bibinfo {year}
  {1978})}\BibitemShut {NoStop}%
\bibitem [{\citenamefont {Ji}\ \emph {et~al.}(2018)\citenamefont {Ji},
  \citenamefont {Zhao}, \citenamefont {Shyy},\ and\ \citenamefont
  {Xu}}]{Ji2018}%
  \BibitemOpen
  \bibfield  {author} {\bibinfo {author} {\bibfnamefont {X.}~\bibnamefont
  {Ji}}, \bibinfo {author} {\bibfnamefont {F.}~\bibnamefont {Zhao}}, \bibinfo
  {author} {\bibfnamefont {W.}~\bibnamefont {Shyy}}, \ and\ \bibinfo {author}
  {\bibfnamefont {K.}~\bibnamefont {Xu}},\ }\href {\doibase
  10.1016/j.jcp.2017.11.036} {\bibfield  {journal} {\bibinfo  {journal}
  {Journal of Computational Physics}\ } (\bibinfo {year} {2018}),\
  10.1016/j.jcp.2017.11.036},\ \Eprint {http://arxiv.org/abs/1707.09921}
  {arXiv:1707.09921} \BibitemShut {NoStop}%
\bibitem [{\citenamefont {Saadat}\ \emph {et~al.}(2020)\citenamefont {Saadat},
  \citenamefont {B\"osch},\ and\ \citenamefont {Karlin}}]{Saadat2020}%
  \BibitemOpen
  \bibfield  {author} {\bibinfo {author} {\bibfnamefont {M.~H.}\ \bibnamefont
  {Saadat}}, \bibinfo {author} {\bibfnamefont {F.}~\bibnamefont {B\"osch}}, \
  and\ \bibinfo {author} {\bibfnamefont {I.~V.}\ \bibnamefont {Karlin}},\
  }\href {\doibase 10.1103/PhysRevE.101.023311} {\bibfield  {journal} {\bibinfo
   {journal} {Phys. Rev. E}\ }\textbf {\bibinfo {volume} {101}},\ \bibinfo
  {pages} {023311} (\bibinfo {year} {2020})}\BibitemShut {NoStop}%
\bibitem [{\citenamefont {Lax}\ and\ \citenamefont {Liu}(1998)}]{Lax1998}%
  \BibitemOpen
  \bibfield  {author} {\bibinfo {author} {\bibfnamefont {P.}~\bibnamefont
  {Lax}}\ and\ \bibinfo {author} {\bibfnamefont {X.}~\bibnamefont {Liu}},\
  }\href {\doibase 10.1137/S1064827595291819} {\bibfield  {journal} {\bibinfo
  {journal} {SIAM Journal on Scientific Computing}\ }\textbf {\bibinfo {volume}
  {19}},\ \bibinfo {pages} {319} (\bibinfo {year} {1998})}\BibitemShut
  {NoStop}%
\bibitem [{\citenamefont {Kurganov}\ and\ \citenamefont
  {Tadmor}(2002)}]{Kurganov2002}%
  \BibitemOpen
  \bibfield  {author} {\bibinfo {author} {\bibfnamefont {A.}~\bibnamefont
  {Kurganov}}\ and\ \bibinfo {author} {\bibfnamefont {E.}~\bibnamefont
  {Tadmor}},\ }\href {\doibase 10.1002/num.10025} {\bibfield  {journal}
  {\bibinfo  {journal} {Numer. Methods Partial Differ. Equ.}\ } (\bibinfo
  {year} {2002}),\ 10.1002/num.10025}\BibitemShut {NoStop}%
\bibitem [{\citenamefont {Inoue}\ and\ \citenamefont
  {Hattori}(1999)}]{Inoue1999}%
  \BibitemOpen
  \bibfield  {author} {\bibinfo {author} {\bibfnamefont {O.}~\bibnamefont
  {Inoue}}\ and\ \bibinfo {author} {\bibfnamefont {Y.}~\bibnamefont
  {Hattori}},\ }\href {\doibase 10.1017/S0022112098003565} {\bibfield
  {journal} {\bibinfo  {journal} {J. Fluid Mech.}\ }\textbf {\bibinfo {volume}
  {380}},\ \bibinfo {pages} {81} (\bibinfo {year} {1999})}\BibitemShut
  {NoStop}%
\bibitem [{\citenamefont {Sterling}\ and\ \citenamefont
  {Chen}(1996)}]{Sterling1996}%
  \BibitemOpen
  \bibfield  {author} {\bibinfo {author} {\bibfnamefont {J.~D.}\ \bibnamefont
  {Sterling}}\ and\ \bibinfo {author} {\bibfnamefont {S.}~\bibnamefont
  {Chen}},\ }\href {\doibase 10.1006/jcph.1996.0016} {\bibfield  {journal}
  {\bibinfo  {journal} {J. Comput. Phys.}\ }\textbf {\bibinfo {volume} {123}},\
  \bibinfo {pages} {196} (\bibinfo {year} {1996})},\ \Eprint
  {http://arxiv.org/abs/9306001} {arXiv:9306001 [comp-gas]} \BibitemShut
  {NoStop}%
\bibitem [{\citenamefont {Hosseini}\ \emph
  {et~al.}(2019{\natexlab{a}})\citenamefont {Hosseini}, \citenamefont
  {Coreixas}, \citenamefont {Darabiha},\ and\ \citenamefont
  {Th\'evenin}}]{Hosseini2019a}%
  \BibitemOpen
  \bibfield  {author} {\bibinfo {author} {\bibfnamefont {S.~A.}\ \bibnamefont
  {Hosseini}}, \bibinfo {author} {\bibfnamefont {C.}~\bibnamefont {Coreixas}},
  \bibinfo {author} {\bibfnamefont {N.}~\bibnamefont {Darabiha}}, \ and\
  \bibinfo {author} {\bibfnamefont {D.}~\bibnamefont {Th\'evenin}},\ }\href
  {\doibase 10.1103/PhysRevE.99.063305} {\bibfield  {journal} {\bibinfo
  {journal} {Phys. Rev. E}\ }\textbf {\bibinfo {volume} {99}},\ \bibinfo
  {pages} {063305} (\bibinfo {year} {2019}{\natexlab{a}})}\BibitemShut
  {NoStop}%
\bibitem [{\citenamefont {Hosseini}\ \emph
  {et~al.}(2019{\natexlab{b}})\citenamefont {Hosseini}, \citenamefont
  {Coreixas}, \citenamefont {Darabiha},\ and\ \citenamefont
  {Th\'evenin}}]{Hosseini2019}%
  \BibitemOpen
  \bibfield  {author} {\bibinfo {author} {\bibfnamefont {S.~A.}\ \bibnamefont
  {Hosseini}}, \bibinfo {author} {\bibfnamefont {C.}~\bibnamefont {Coreixas}},
  \bibinfo {author} {\bibfnamefont {N.}~\bibnamefont {Darabiha}}, \ and\
  \bibinfo {author} {\bibfnamefont {D.}~\bibnamefont {Th\'evenin}},\ }\href
  {\doibase 10.1103/PhysRevE.100.063301} {\bibfield  {journal} {\bibinfo
  {journal} {Phys. Rev. E}\ }\textbf {\bibinfo {volume} {100}},\ \bibinfo
  {pages} {063301} (\bibinfo {year} {2019}{\natexlab{b}})}\BibitemShut
  {NoStop}%
\bibitem [{\citenamefont {White}\ and\ \citenamefont
  {Chong}(2011)}]{White2011}%
  \BibitemOpen
  \bibfield  {author} {\bibinfo {author} {\bibfnamefont {A.~T.}\ \bibnamefont
  {White}}\ and\ \bibinfo {author} {\bibfnamefont {C.~K.}\ \bibnamefont
  {Chong}},\ }\href {\doibase 10.1016/j.jcp.2011.04.031 M4 - Citavi} {\bibfield
   {journal} {\bibinfo  {journal} {J. Comput. Phys.}\ }\textbf {\bibinfo
  {volume} {230}},\ \bibinfo {pages} {6367} (\bibinfo {year}
  {2011})}\BibitemShut {NoStop}%
\bibitem [{\citenamefont {Watari}\ and\ \citenamefont
  {Tsutahara}(2006)}]{Watari2006}%
  \BibitemOpen
  \bibfield  {author} {\bibinfo {author} {\bibfnamefont {M.}~\bibnamefont
  {Watari}}\ and\ \bibinfo {author} {\bibfnamefont {M.}~\bibnamefont
  {Tsutahara}},\ }\href {\doibase 10.1016/j.physa.2005.06.103} {\bibfield
  {journal} {\bibinfo  {journal} {Phys. A Stat. Mech. its Appl.}\ }\textbf
  {\bibinfo {volume} {364}},\ \bibinfo {pages} {129} (\bibinfo {year}
  {2006})}\BibitemShut {NoStop}%
\end{thebibliography}%

\end{document}